\documentclass[journal=jctcce,manuscript=article,layout=twocolumn]{achemso}

\usepackage[colorlinks,linkcolor=blue,citecolor=blue,urlcolor=black,bookmarks=false,hypertexnames=true]{hyperref} 

\usepackage[version=3]{mhchem}
\usepackage{amssymb}
\usepackage{amsmath}
\usepackage[utf8]{inputenc}
\usepackage{textcomp}
\usepackage{color}
\usepackage{braket}
\usepackage{xspace}
\usepackage{cleveref}
\usepackage{graphicx}
\usepackage{subfigure}
\usepackage{threeparttable}
\usepackage{textcomp}
\usepackage{setspace}
\usepackage{caption}
\usepackage{comment}

\usepackage{array}
\newcolumntype{L}[1]{>{\raggedright\let\newline\\\arraybackslash\hspace{0pt}}m{#1}}
\newcolumntype{C}[1]{>{\centering\let\newline\\\arraybackslash\hspace{0pt}}m{#1}}
\newcolumntype{R}[1]{>{\raggedleft\let\newline\\\arraybackslash\hspace{0pt}}m{#1}}

\makeatletter
\let\l@addto@macro\relax
\makeatother
\usepackage[fontsize=11pt]{scrextend}

\let\oldmaketitle\maketitle
\let\maketitle\relax

\newcommand{\cm}{\ensuremath{\text{cm}^{-1}}\xspace}

\newcommand*{\mae}{\ensuremath{{\mathrm{MAE}}}\xspace}

\DeclareUnicodeCharacter{2009}{\,}

\crefname{figure}{Figure}{Figures}
\crefname{table}{Table}{Tables}
\crefname{equation}{Eq.}{Eqs.}
\crefname{section}{Section}{Sections}
\crefname{subsection}{Section}{Sections}

\author{Rajat~Majumder$^\dag$}
\affiliation{$^\dag$Department of Chemistry and Biochemistry, The Ohio State University, Columbus, Ohio 43210, USA}

\author{Alexander~Yu.~Sokolov$^\dag$}
\email{sokolov.8@osu.edu}
\affiliation{$^\dag$Department of Chemistry and Biochemistry, The Ohio State University, Columbus, Ohio 43210, USA}

\begin{tocentry}
\includegraphics[width=1.0\textwidth]{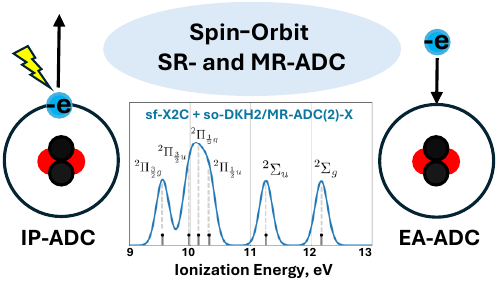}
\end{tocentry}

\title{{\color{blue}
    Algebraic Diagrammatic Construction Theory of Charged Excitations With Consistent Treatment of Spin--Orbit Coupling and Dynamic Correlation
}}

\begin{document}


\newcommand*{\abstractext}{
We present algebraic diagrammatic construction theory for simulating spin--orbit coupling and electron correlation in charged electronic states and photoelectron spectra.
Our implementation supports Hartree--Fock and multiconfigurational reference wavefunctions, enabling efficient correlated calculations of relativistic effects using single-reference (SR-) and multireference (MR-) ADC.
We combine the SR- and MR-ADC methods with three flavors of spin--orbit two-component Hamiltonians and benchmark their performance for a variety of atoms and small molecules.
When multireference effects are not important, the SR-ADC approximations are competitive in accuracy to MR-ADC, often showing closer agreement with experimental results. 
However, for electronic states with multiconfigurational character and in non-equilibrium regions of potential energy surfaces, the MR-ADC methods are more reliable, predicting accurate excitation energies and zero-field splittings.
Our results demonstrate that the spin--orbit ADC methods are promising approaches for interpreting and predicting the results of modern spectroscopies. 
\vspace{0.25cm}
}

\twocolumn[
\begin{@twocolumnfalse}
\oldmaketitle
\vspace{-0.75cm}
\begin{abstract}
\abstractext
\end{abstract}
\end{@twocolumnfalse}
]

\section{Introduction}
\label{sec:introduction}
Charged excitations are perturbations to a chemical system that result in the net change of electron number and charge state.
Detailed understanding of these processes is crucial to advancing several key areas, such as developing better photoredox catalysts and semiconductor materials\cite{Morab2023:p1657,Sur2023:p100190,Koike2023:p100205}, improving atmospheric and combustion models\cite{Umstead:1966p293,Uddin:2020p1905739}, and characterizing radiation damage in biomolecules.\cite{Steenken:1989p503,Yan:1992p1983,Huels:1998p1309}
Charged excitations are also the primary electronic transitions studied in photoelectron spectroscopy that uses high-energy light (UV, XUV, or X-ray) to measure electron binding energies.\cite{Tonti2009:p673695,Marsh2018:p54502,Perry2021:p29902996}
Recent developments in time-resolved photoelectron spectroscopy enable probing the dynamics of charged electronic states and emitted electrons with atto- and femtosecond time resolution.\cite{Zanni1999:p37483755,Stolow2004:p17191757,Zhang2009:p123601,Jadoun2021:p81038108}

Understanding the electronic structure and dynamics of charged excited states requires insights from accurate theoretical calculations. 
However, simulating charged excitations faces many difficulties associated with the description of orbital relaxation, charge localization, and electronic spin.
To accurately capture these properties, a variety of electronic structure methods that incorporate electron correlation starting with a single- or multireference wavefunction are available.
These approaches range from lower-cost response
\cite{Runge:1984p997,Besley2010:p12024112039,Akama2010:p54104,Hedin1965:pa796,Nakatsuji1978:p20532065,Nakatsuji1979:p329333,Lopata:2011ce,Mckechnie:2015p194114,Faleev:2004p126406,vanSchilfgaarde:2006p226402,Reining:2017pe1344,Linderberg:2004} 
and perturbation theories
\cite{Binkley1975:229236,Bartlett:1981p359,Pulay1986:p357368,Knowles1991:p130136,Hirao1992:p374380,Murphy1992:p41704184,Zaitsevskii:1995p597,Andersson1998:p1218,Angeli:2001p10252,Ghigo:2004p142,Shavitt:2008p5711,Granovsky:2011p214113,Chen:2014ji,Sharma:2016p034103,Sokolov:2016p064102} to more computationally expensive and accurate configuration interaction
\cite{Lischka1981:p91100,Knowles1984:p315321,Sherrill1999:p143269,Werner:2008hj,Shamasundar2011:p054101}
and coupled cluster methods.
\cite{Stanton1993:p70297039,Chattopadhyay:2000p7939,Nayak:2006fh,Crawford2007:p33136,Shavitt_Bartlett_2009,Datta:2011p214116,Evangelista:2011p224102,Evangelista:2011p114102,Henderson:2014jw,Eriksen:2015dd,Garniron:2017wt} 

In addition to electron correlation, simulating charged excitations may require taking into account spin--orbit coupling. 
Along with scalar relativistic effects, spin--orbit interactions are important for excitations from core $p$- and $d$-orbitals and are critical to the electronic structure of molecules with heavy elements.
Accurate treatment of electron correlation and relativistic effects can be achieved using four-component theories based on the Dirac--Coulomb (DC) or Dirac--Coulomb--Breit (DCB) Hamiltonians.\cite{Yanai:2001p65266538,Schwarz2010:p162,Fleig:2012p2,Pyykko:2012p45}
However, the computational costs of four-component methods are significantly higher than those of nonrelativistic electronic structure theories, limiting the scope of their applications.

A more economical strategy to simultaneously capture electron correlation and spin--orbit coupling is offered by the two-component relativistic theories.
These approaches are formulated by decoupling the electronic and positronic states in the Dirac equation  and using the resulting two-component Hamiltonian to describe electron correlation.
Two-component methods can be broadly divided into two classes: (i) variational, which introduce spin--orbit interactions in the reference wavefunction,\cite{Hess:1986p3742,Jensen:1996p40834097,Fleig2001:p47754790,Liu:2010p1679,Saue:2011p3077,Fleig:2012p2,Pyykko:2012p45,Reiher:2004p10945,Kutzelnigg:2012p16,Ganyushin:2013p104113,Hu:2020p29752984} or (ii) perturbative, which first calculate a spin-free relativistic reference wavefunction and incorporate dynamic correlation with spin--orbit coupling {\it a posteriori}.\cite{Vallet2000:p13911402,finley:2001p042502,Roos:2004p29192927,Kleinschmidt:2006p124101,Mai2014:p074105,Cheng:2014p164107,Meitei:2020p3597} 
Most perturbative two-component theories treat spin--orbit coupling as a first-order perturbation and describe dynamic correlation at a higher level of theory.
While the first-order approximation is accurate for compounds with light elements at low excitation energies, it is unreliable for electronic states with strong relativistic effects.\cite{Majumder2024:p46764688}

In this work, we present an efficient approach for simulating charged excitations that (i) captures static correlation in frontier molecular orbitals, (ii) treats dynamic correlation and spin--orbit coupling as equal perturbations to the nonrelativistic Hamiltonian, and (iii) incorporates their effects in excitation energies and transition intensities up to the second order in perturbation theory.
Our approach is formulated in the framework of multireference algebraic diagrammatic construction theory (MR-ADC)\cite{Sokolov:2018p149,Sokolov2024:p121155} that allows to efficiently simulate neutral and charged excitations by approximating linear response functions using low-order multireference perturbation theory.\cite{Chatterjee2019:p59085924,Chatterjee2020:p63436357,Mazin2021:p61526165,DeMoura2022:p47694784,Mazin2023:p49915006,Gaba2024:p1592715938,DeMoura2024:p58165831} 
Four-component implementations of single-reference ADC (SR-ADC)\cite{Schirmer1982:p23952416,Schirmer1983:p12371259,Schirmer1991:p46474659,Mertins1996:p21402152,Trofimov2006:p110,Dreuw2015p8295,Dempwolff2019:p064108,Dempwolff2020:p024125,Dempwolff2021:p074105,Banerjee2023:p3053,Leitner2022:p184101} with the variational treatment of spin--orbit effects and perturbative description of dynamic correlation in charged\cite{Pernpointner2004:p87828791,Pernpointner2010:p205102,Hangleiter2010:p205102,Pernpointner2018:p15101522} and neutral excitations\cite{Pernpointner2014:p84108,Krauter2017:p286293,Chakraborty2024:arxiv} have been reported.

Here, we implement and benchmark the MR-ADC methods for simulating electron-attached (EA) and ionized (IP) states incorporating dynamic correlation and spin--orbit coupling effects up to the second order in perturbation theory.
The spin--orbit interactions are described using the Breit--Pauli (BP), \cite{Breit:1932p616,Bearpark:1993p479502,Berning:2000p1823} exact two-component first-order Douglas--Kroll--Hess (sf-X2C+so-DKH1), or exact two-component second-order Douglas--Kroll--Hess (sf-X2C+so-DKH2) Hamiltonians\cite{Li:2012p154114,Li:2014p054111,Cao:2017p3713,Wang:2023p848855} within the mean-field spin--orbit approximation.\cite{Hess:1996p365, Berning:2000p1823,Li:2014p054111,Cao:2017p3713,Wang:2023p848855} 
Starting with a single-determinant (Hartree--Fock) reference wavefunction, our  MR-ADC methods reduce to the spin--orbit SR-ADC approximations, for which results are also presented. 

\section{Theory}
\label{sec:theory}

\subsection{Algebraic Diagrammatic Construction Theory of Charged Excitations}
\label{sec:theory:overview}

Algebraic diagrammatic construction (ADC) belongs to a class of propagator theories that describe charged excitations in terms of the one-particle Green's function (1-GF).\cite{Schirmer1982:p23952416,Schirmer1983:p12371259,Schirmer1991:p46474659,Mertins1996:p21402152,Trofimov2006:p110,Dreuw2015p8295,Dempwolff2019:p064108,Dempwolff2020:p024125,Dempwolff2021:p074105,Banerjee2023:p3053,Leitner2022:p184101} 
For the $N$-electron reference electronic state $|{\Psi^{N}}\rangle$ with energy $E_{N}$ (usually, the ground state), 1-GF can be expressed as
\begin{align}
\label{eq:1gf}
G_{pq}(\omega) &= G^{+}_{pq}(\omega) + G^{-}_{pq}(\omega) \notag \\
&= \langle{\Psi^{N}}|a_p (\omega - H + E_{N})^{-1} a^{\dagger}_q |{\Psi^{N}}\rangle \notag \\
&+ \langle{\Psi^{N}}|a^{\dagger}_q (\omega + H - E_{N})^{-1} a_p |{\Psi^{N}}\rangle 
\end{align}
where $G^{+}_{pq}(\omega)$ and $G^{-}_{pq}(\omega)$ are the forward and backward components of 1-GF, $H$ is the electronic Hamiltonian, and $\omega$ is the frequency of radiation promoting the charged excitations. 
The $a^{\dagger}_p$/$a_p$ are the creation/annihilation operators describing electron addition/removal. 
Alternatively, 1-GF can be written in a spectral representation
\begin{align}
\label{eq:spectral_gf}
G_{pq}(\omega) &= \sum_n \frac{\langle{\Psi^{N}}|a_p|{\Psi^{N+1}_n}\rangle \langle{\Psi^{N+1}_n}|a^{\dagger}_q|{\Psi^{N}}\rangle}{\omega - E_{N+1,n} + E_{N}} \notag \\
			&+ \sum_n \frac{\langle{\Psi^{N}}|a^{\dagger}_q|{\Psi^{N-1}_n}\rangle \langle{\Psi^{N-1}_n}|a_p|{\Psi^{N}}\rangle}{\omega + E_{N-1,n} - E_{N}} 
\end{align}
that encodes information about the vertical electron affinities ($E_{N+1,n} - E_{N}$), ionization energies ($E_{N-1,n} - E_{N}$), and the corresponding transition probabilities ($\langle{\Psi^{N}}|a_p|{\Psi^{N+1}_n}\rangle \langle{\Psi^{N+1}_n}|a^{\dagger}_q|{\Psi^{N}}\rangle$ and $\langle{\Psi^{N}}|a^{\dagger}_q|{\Psi^{N-1}_n}\rangle \langle{\Psi^{N-1}_n}|a_p|{\Psi^{N}}\rangle$). 

ADC approximates the exact 1-GF by expressing each term in \cref{eq:spectral_gf} as a product of non-diagonal matrices:
\begin{align}
	\label{eq:spectral_matrix}
	\mathbf{G}_{\pm}(\omega) = \mathbf{T}_{\pm} (\omega \mathbf{S}_{\pm} - \mathbf{M}_{\pm})^{-1} \mathbf{T}_{\pm}
\end{align}
Here, $\mathbf{M}_{\pm}$ and $\mathbf{T}_{\pm}$ are the effective Hamiltonian and transition moments matrices that provide information about vertical charged excitation energies and transition probabilities, respectively.
Each matrix is expressed in a basis of ($N\pm1$)-electron excited-state configurations that are, in general, nonorthogonal with overlap integrals stored in $\mathbf{S}_\pm$.
Approximating $\mathbf{M}_{\pm}$, $\mathbf{T}_{\pm}$, and $\mathbf{S}_\pm$ using perturbation theory up to the order $n$
\begin{align}
	\label{eq:pertM}
	\mathbf{M}_{\pm} &\approx \mathbf{M}^{(0)}_{\pm} + \mathbf{M}^{(1)}_{\pm} + \dotsc + \mathbf{M}^{(n)}_{\pm} \\
	\label{eq:pertT}
	\mathbf{T}_{\pm} &\approx \mathbf{T}^{(0)}_{\pm} + \mathbf{T}^{(1)}_{\pm} + \dotsc + \mathbf{T}^{(n)}_{\pm} \\
	\label{eq:pertS}
	\mathbf{S}_{\pm} &\approx \mathbf{S}^{(0)}_{\pm} + \mathbf{S}^{(1)}_{\pm} + \dotsc + \mathbf{S}^{(n)}_{\pm} 
\end{align}
defines the $n$th-order ADC approximation (ADC($n$)). 

Diagonalizing the $\mathbf{M}_{\pm}$ matrices allows to compute charged excitation energies ($\mathbf{\Omega}_{\pm}$):
\begin{align}
	\mathbf{M}_{\pm}\mathbf{Y}_{\pm} = \mathbf{S}_{\pm}\mathbf{Y}_{\pm}\mathbf{\Omega}_{\pm}
\end{align}
The corresponding eigenvectors $\mathbf{Y}_{\pm}$ can be combined with the transition moments matrices $\mathbf{T}_{\pm}$ to compute spectroscopic amplitudes
\begin{align}
	\mathbf{X}_{\pm} = \mathbf{T}_{\pm} \mathbf{S}^{-1/2}_{\pm}\mathbf{Y}_{\pm}
\end{align}
which provide information about the probabilities of charged excitations. 

\subsection{Multireference ADC}
\label{sec:theory:nonrel_mradc}

\begin{figure*}[t!]
	\includegraphics[width=0.6\textwidth]{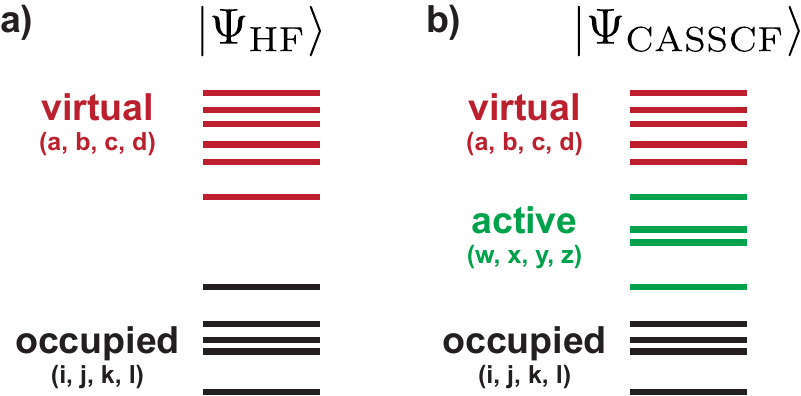} 
	\captionsetup{justification=justified,singlelinecheck=false,font=footnotesize}
	\caption{
		Schematic diagram representing molecular orbitals and their labels for 
		a) the Hartree--Fock (HF) reference wavefunction in SR-ADC and 
		(b) the CASSCF reference wavefunction in MR-ADC. 
		Reproduced from Ref.\@\citenum{Banerjee2023:p3053} with permission from the American Chemical Society. 
		Copyright 2023.
	} 
	\label{fig:orbital_diagram}
\end{figure*}

Two ADC formulations have been proposed: single-reference (SR-)\cite{Schirmer1982:p23952416,Schirmer1983:p12371259,Schirmer1991:p46474659,Mertins1996:p21402152,Trofimov2005:p144115,Trofimov2006:p110,Dreuw2015p8295,Dempwolff2019:p064108,Dempwolff2020:p024125,Dempwolff2021:p074105,Banerjee2019:p224112,Banerjee2023:p3053,Leitner2022:p184101}  and multireference (MR-)\cite{Sokolov:2018p149,Chatterjee2019:p59085924,Chatterjee2020:p63436357,Mazin2021:p61526165,DeMoura2022:p47694784,Mazin2023:p49915006,Gaba2024:p1592715938,DeMoura2024:p58165831,Sokolov2024:p121155} ADC.
In SR-ADC, contributions to $\mathbf{M}_{\pm}$, $\mathbf{T}_{\pm}$, and $\mathbf{S}_\pm$ are evaluated using M\o ller--Plesset perturbation theory\cite{Cremer2011:p509530} following a Hartree--Fock calculation for the reference state (\cref{fig:orbital_diagram}a).
MR-ADC starts with a complete active space self-consistent field (CASSCF, \cref{fig:orbital_diagram}b) reference wavefunction $\ket{\Psi_0}$ and incorporates dynamic correlation effects using multireference $N$-electron valence perturbation theory.\cite{Angeli:2001p10252,Angeli:2002p9138,Angeli:2006p054108}
If the number of active orbitals in the CASSCF reference wavefunction is zero, the MR-ADC($n$) methods reduce to the SR-ADC($n$) approximations.

Perturbative contributions to the MR-ADC($n$) matrices in \cref{eq:pertM,eq:pertT,eq:pertS} can be expressed as:\cite{Chatterjee2019:p59085924,Chatterjee2020:p63436357}
\begin{align}
	\label{eq:Mmatrix_ea}
	M^{(n)}_{+\mu\nu} &= \sum^{k+l+m=n}_{klm} \langle{\Psi_0}|[h^{(k)}_{+\mu} , [\tilde{H}^{(l)}, h^{(m)\dagger}_{+\nu} ]]_{+}|{\Psi_0}\rangle \\
	\label{eq:Tmatrix_ea}
	T^{(n)}_{+p\nu} &= \sum^{k+l=n}_{kl} \langle{\Psi_0}|[\tilde{a}^{(k)}_p , h^{(l)\dagger}_{+\nu}]_+|{\Psi_0}\rangle \\
	\label{eq:Smatrix_ea}
	S^{(n)}_{+\mu\nu} &= \sum^{k+l=n}_{kl} \langle{\Psi_0}|[h^{(k)}_{+\mu} , h^{(l)\dagger}_{+\nu}]_+|{\Psi_0}\rangle \\
	\label{eq:Mmatrix_ip}
	M^{(n)}_{-\mu\nu} &= \sum^{k+l+m=n}_{klm} \langle{\Psi_0}|[h^{(k)\dagger}_{-\mu} , [\tilde{H}^{(l)}, h^{(m)}_{-\nu} ]]_{+}|{\Psi_0}\rangle \\
	\label{eq:Tmatrix_ip}
	T^{(n)}_{-p\nu} &= \sum^{k+l=n}_{kl} \langle{\Psi_0}|[\tilde{a}^{(k)}_p , h^{(l)}_{-\nu}]_+|{\Psi_0}\rangle \\
	\label{eq:Smatrix_ip}
	S^{(n)}_{-\mu\nu} &= \sum^{k+l=n}_{kl} \langle{\Psi_0}|[h^{(k)\dagger}_{-\mu} , h^{(l)}_{-\nu}]_+|{\Psi_0}\rangle 
\end{align} 
Here, $[A,B] = AB-BA$ denotes a commutator, $[A,B]_+ = AB+BA$ is an anticommutator, while $\tilde{H}^{(k)}$, $\tilde{a}^{(k)}_p$, and $h^{(k)\dagger}_{\pm\nu}$ are the $k$th-order contributions to effective Hamiltonian ($\tilde{H}$), effective observable ($\tilde{a}_p$), and excitation manifold ($h^{\dagger}_{\pm\nu}$) operators, respectively. 

The low-order $\tilde{H}^{(k)}$ and $\tilde{a}^{(k)}_p$ have the form:\cite{Chatterjee2019:p59085924,Chatterjee2020:p63436357}
\begin{align}
	\label{eq:eff_H0}
	\tilde{H}^{(0)} &= H^{(0)} \\
	\label{eq:eff_H1}
	\tilde{H}^{(1)} &= V + [H^{(0)}, T^{(1)} - T^{(1)\dagger}] \\
	\label{eq:eff_H2}
	\tilde{H}^{(2)} &= [H^{(0)}, T^{(2)} - T^{(2)\dagger}] + \frac{1}{2} [V + \tilde{H}^{(1)}, T^{(1)} - T^{(1)\dagger}] \\
	\label{eq:eff_aop0}
	\tilde{a}^{(0)}_{p} &= a_p \\
	\label{eq:eff_aop1}
	\tilde{a}^{(1)}_{p} &= [a_{p}, T^{(1)} - T^{(1)\dagger}] \\
	\label{eq:eff_aop2}
	\tilde{a}^{(2)}_{p} &= [a_{p}, T^{(2)} - T^{(2)\dagger}] + \frac{1}{2} [[a_{p}, T^{(1)} - T^{(1)\dagger}], T^{(1)} - T^{(1)\dagger}]
\end{align}
where $H^{(0)}$ is the Dyall zeroth-order Hamiltonian,\cite{Dyall:1998p4909} $V = H - H^{(0)}$ is the perturbation operator, and $T^{(k)}$ is the $k$th-order cluster correlation operator. 
The Dyall Hamiltonian $H^{(0)}$ incorporates the one- and two-electron active-space terms of the electronic Hamiltonian $H$ and describes the static electron correlation in active orbitals.\cite{Sokolov2024:p121155}
The $V$ and $T^{(k)}$ operators incorporate dynamic correlation in non-active orbitals. 
Up to the second order in multireference perturbation theory, $T^{(k)}$ ($k \le 2$) incorporates single and double excitations out of the reference wavefunction $|{\Psi_0}\rangle$ and can be written as
\begin{align}
	\label{eq:amp_form}
	T^{(k)} &= \sum_{\mu} t^{(k)}_{\mu} \tau^{\dagger}_{\mu}
\end{align}
where the amplitudes $t^{(k)}_{\mu} $ are determined by projecting the $k$th-order effective Hamiltonian on the singly and doubly excited configurations $\tau_\mu^\dag|{\Psi_0}\rangle$:\cite{Chatterjee2019:p59085924,Chatterjee2020:p63436357}
\begin{align}
	\label{eq:amp_eq}
	\langle{\Psi_0}|\tau_\mu \tilde{H}^{(k)}|{\Psi_0}\rangle = 0
\end{align}

\begin{figure*}[t!]
    \includegraphics[width=0.9\textwidth]{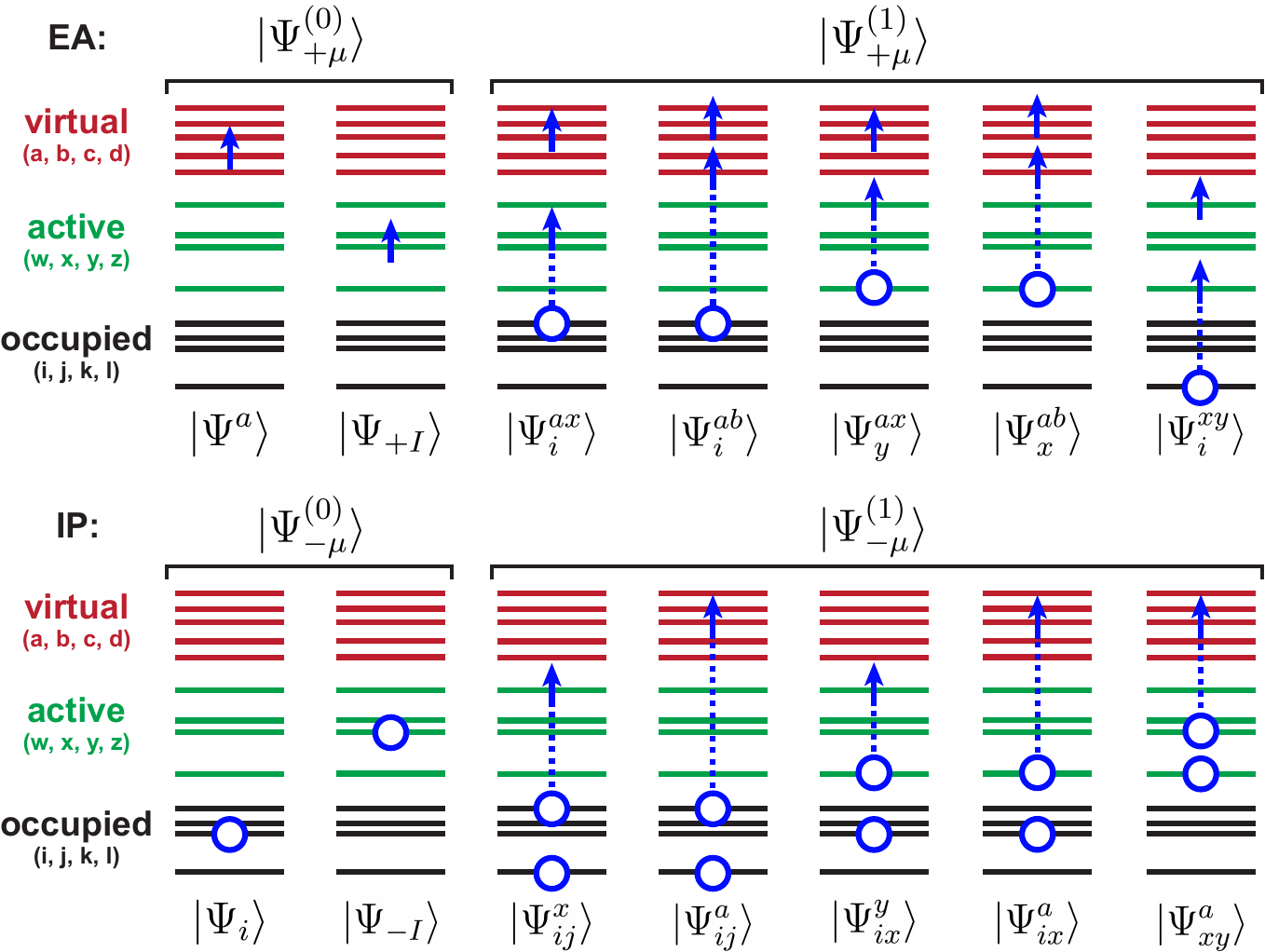} 
    \captionsetup{justification=raggedright,singlelinecheck=false}
	\caption{Schematic illustration of the electron-attached and ionized states produced by acting the $h^{(k)\dagger}_{\pm \mu}$ ($k$ = 0, 1) operators on the reference state $|{\Psi_0}\rangle$ in MR-ADC(2) and MR-ADC(2)-X. 
		An arrow represents electron attachment, a circle denotes ionization, and a circle connected with an arrow indicates single excitation. 
		The states $\ket{\Psi_{\pm I}}$ incorporate all $(N\pm1)$-electron excitations in the active orbitals. 
		Reproduced from Ref.\@\citenum{Banerjee2023:p3053} with permission from the American Chemical Society. 
		Copyright 2023.
		}
	\label{fig:excitations}
\end{figure*}

Finally, the excitation manifold operators $h^{(k)\dagger}_{\pm\nu}$ are used to represent $\tilde{H}^{(k)}$ and $\tilde{a}^{(k)}_p$ in \cref{eq:Mmatrix_ea,eq:Tmatrix_ea,eq:Smatrix_ea,eq:Mmatrix_ip,eq:Tmatrix_ip,eq:Smatrix_ea} in the basis of $(N\pm1)$-electron  electronic configurations ($h^{(k)\dagger}_{\pm\nu}\ket{\Psi_0}$).\cite{Chatterjee2019:p59085924,Chatterjee2020:p63436357}
These multireference wavefunctions are depicted in \cref{fig:excitations} for $k$ = 0 and 1.
The $h^{(0)\dagger}_{\pm\nu}$ operators incorporate all $(N\pm1)$-electron excitations in the active space ($\ket{\Psi_{\pm I}}$) and the one-electron attachment/ionization in virtual/core orbitals ($\ket{\Psi^a}$/$\ket{\Psi_i}$), respectively. 
The charged excitations out of active space involving two electrons are described by $h^{(1)\dagger}_{\pm\nu}$.

\begin{figure*}[t!]
	\includegraphics[width=0.9\textwidth]{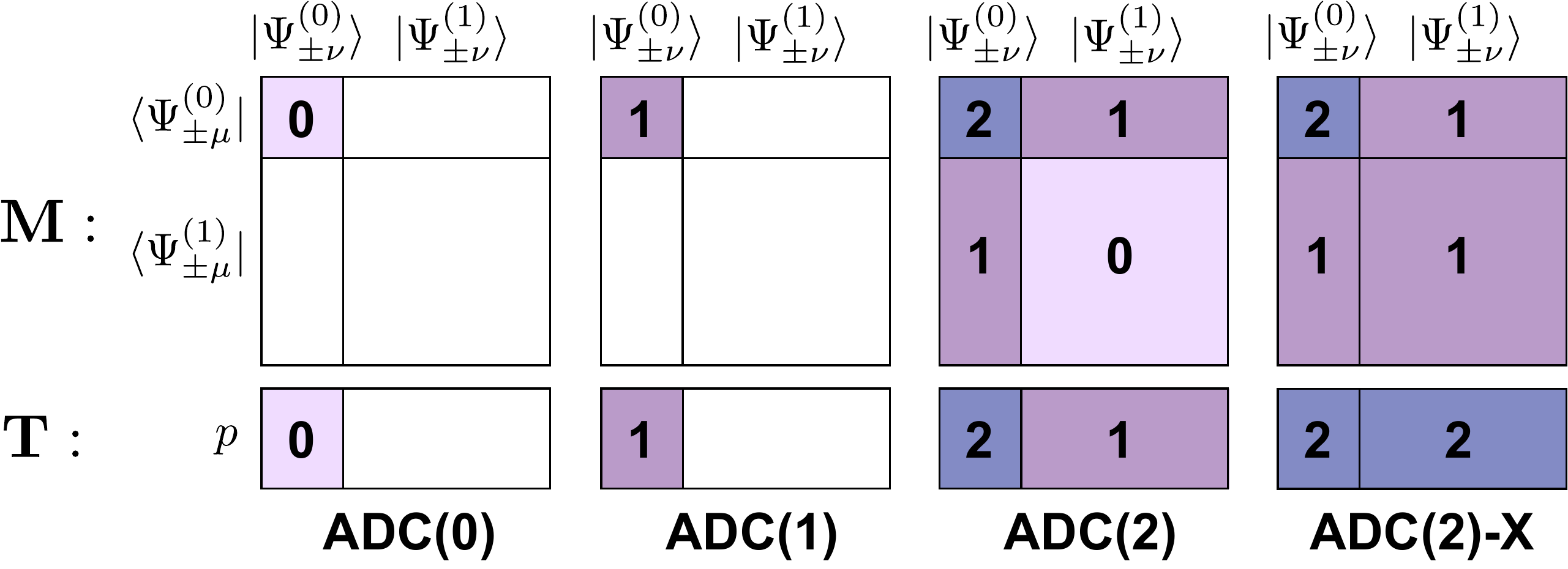} 
	\captionsetup{justification=justified,singlelinecheck=false,font=footnotesize}
	\caption{
		Perturbative structures of the effective Hamiltonian ($\mathbf{M}_\pm$) and transition moments ($\mathbf{T}_\pm$) matrices in the low-order MR-ADC approximations. 
		Numbers denote the perturbation order to which the effective Hamiltonian and transition moments are expanded for each sector. 
		Shaded areas indicate nonzero blocks. 
		Adapted from Ref.\@\citenum{Banerjee2023:p3053} with permission from the American Chemical Society. 
		Copyright 2023.
	} 
	\label{fig:mradc_matrices}
\end{figure*}

\cref{eq:Mmatrix_ea,eq:Tmatrix_ea,eq:Smatrix_ea,eq:Mmatrix_ip,eq:Tmatrix_ip,eq:Smatrix_ea} define the perturbative structure of MR-ADC($n$) matrices where the sum of orders for $h^{(k)\dagger}_{\pm\nu}$, $\tilde{H}^{(l)}$, and $\tilde{a}^{(m)}_p$  cannot exceed $n$ for a particular matrix element. 
\cref{fig:mradc_matrices} illustrates this for the low-order MR-ADC methods. 
In addition to the strict MR-ADC(0), MR-ADC(1), and MR-ADC(2) approximations, an extended second-order MR-ADC method (MR-ADC(2)-X) has been developed, which incorporates higher-order terms in $\mathbf{M}_{\pm}$ and $\mathbf{T}_{\pm}$ for the description of double excitations ($h^{(1)\dagger}_{\pm\nu}$).\cite{Chatterjee2020:p63436357}
These additional terms provide a higher-order description of orbital relaxation effects in excited states.
Keeping the size of active space constant, MR-ADC(2) and MR-ADC(2)-X have the $\mathcal{O}(N^5)$ computational scaling with the basis set size ($N$), which allows to perform calculations for molecules with more than 1000 molecular orbitals.\cite{DeMoura2024:p58165831}

\subsection{Incorporating Relativistic Effects in MR-ADC}
\label{sec:theory:soc_mradc}

The goal of this work is to incorporate relativistic effects in the MR-ADC calculations of charged electronic states without significantly increasing their computational cost.
To achieve this, we employ three variants of two-component relativistic Hamiltonians, namely: 
i) Breit--Pauli (BP),\cite{Breit:1932p616,Bearpark:1993p479502,Berning:2000p1823} 
ii) exact two-component first-order Douglas--Kroll--Hess (sf-X2C+so-DKH1),\cite{Li:2012p154114,Li:2014p054111} 
and iii) exact two-component second-order Douglas--Kroll--Hess (sf-X2C+so-DKH2).\cite{Li:2014p054111}
These Hamiltonians are derived by approximately decoupling the electronic and positronic degrees of freedom in the four-component Dirac equation and subsequently adding the Coulomb and Gaunt two-electron terms.
The BP Hamiltonian represents the lowest level of decoupling, which is valid when relativistic effects are weak but is variationally unstable (not bounded from below) and becomes increasingly inaccurate as relativistic effects get stronger. 
The sf-X2C+so-DKH1 and sf-X2C+so-DKH2 Hamiltonians used in this work are formulated using the spin-free exact two-component approach of Liu and co-workers (X2C-1e),\cite{Li:2012p154114} which provides a more accurate description of scalar relativistic terms than the conventional DKH1 and DKH2 Hamiltonians.\cite{Douglas:1974p89,Hess:1986p3742,Jansen:1989p6016}
We refer the readers to excellent reviews on this topic for additional information.\cite{Liu:2010p1679,Fleig:2012p2,Liu2020:p180901}

Each two-component Hamiltonian can be expressed in a general form as:
\begin{align}
	\label{eq:h_2c_general}
	 {H}_{\mathrm{2c}} &= {H}_{\mathrm{SF}} + H_{\mathrm{SO}}
\end{align}
where ${H}_{\mathrm{SF}}$ describes the scalar relativistic effects and $H_{\mathrm{SO}}$ incorporates spin--orbit coupling.
For BP and sf-X2C+so-DKH1, we choose ${H}_{\mathrm{SF}}$ to be the X2C-1e Hamiltonian\cite{Li:2012p154114} that captures the scalar relativistic effects more accurately than the spin-free contributions of the conventional BP and DKH1 Hamiltonians (${H}_{\mathrm{SF}} = {H}^{\mathrm{X2C-1e}}_{\mathrm{SF}}$). 
For sf-X2C+so-DKH2, ${H}_{\mathrm{SF}}$ is defined as the X2C-1e Hamiltonian plus additional terms from the second-order DKH transformation due to the picture change effect (${H}_{\mathrm{SF}} = {H}^{\mathrm{X2C-1e}}_{\mathrm{SF}} + {H}^{\mathrm{DKH2}}_{\mathrm{SF}}$). 
Working equations for ${H}^{\mathrm{X2C-1e}}_{\mathrm{SF}}$ and ${H}^{\mathrm{DKH2}}_{\mathrm{SF}}$ can be found in Ref.\@ \citenum{Li:2014p054111}.
For brevity, we will refer to sf-X2C+so-DKH$n$ as DKH$n$ ($n$ = 1, 2) henceforth.

Within the spin--orbit mean-field approximation (SOMF),\cite{Hess:1996p365, Berning:2000p1823} the BP, DKH1, and DKH2 spin-dependent Hamiltonians can be written in a general form:\cite{Bearpark:1993p479502,Berning:2000p1823,Li:2012p154114,Li:2014p054111}
\begin{align}
	\label{eq:h_2c_so_general}
	{H}_{\mathrm{SO}} &= i \frac{\alpha^2}{4} \sum_{\xi} \sum_{pq} F^{\xi}_{pq} {D}^{\xi}_{pq} 
\end{align}
where $\alpha = 1 / c$ is the fine-structure constant, the indices $(p,q,\ldots)$ label all spatial molecular orbitals in the one-electron basis set, $\xi=x,y,z$ denotes Cartesian coordinates, and ${D}^{\xi}_{pq}$ are the one-electron spin excitation operators
\begin{align}
	{D}^{x}_{pq} &= {a}^{\dagger}_{p\alpha}{a}_{q\beta} + {a}^{\dagger}_{p\beta}{a}_{q\alpha}  \\
	{D}^{y}_{pq} &= i ({a}^{\dagger}_{p\beta}{a}_{q\alpha} - {a}^{\dagger}_{p\alpha}{a}_{q\beta})  \\
	{D}^{z}_{pq} &= {a}^{\dagger}_{p\alpha}{a}_{q\alpha} - {a}^{\dagger}_{p\beta}{a}_{q\beta} 
\end{align}
with $\alpha$ and $\beta$ denoting the spin-up and spin-down electrons, respectively. 
The expressions for the matrix elements $F^{\xi}_{pq}$ of each two-component Hamiltonian can be found in Ref.\@ \citenum{Majumder2024:p46764688}.

In our formulation of spin--orbit MR-ADC, we incorporate the scalar relativistic effects in the reference CASSCF calculation by including ${H}_{\mathrm{SF}}$ in the zeroth-order Hamiltonian 
\begin{align}
	H^{(0)}_{\mathrm{2c}} = H^{(0)} + {H}_{\mathrm{SF}}
\end{align}
To describe spin--orbit coupling, we define a new perturbation operator
\begin{align}
	\label{eq:v_2c}
	V_{\mathrm{2c}} 
	= V + {H}_{\mathrm{SO}}
	= H - H^{(0)} + {H}_{\mathrm{SO}}
\end{align}
where $V$ captures dynamic correlation in non-active orbitals (\cref{sec:theory:nonrel_mradc}) and the two component spin--orbit operator ${H}_{\mathrm{SO}}$ is defined in \cref{eq:h_2c_so_general}.
Replacing $H^{(0)}$ by $H^{(0)}_{\mathrm{2c}}$ and $V$ by $V_{\mathrm{2c}}$ in \cref{eq:eff_H0,eq:eff_H1,eq:eff_H2,eq:eff_aop0,eq:eff_aop1,eq:eff_aop2} allows to formulate the MR-ADC($n$) methods with consistent perturbative treatment of dynamic correlation and spin--orbit coupling effects. 

Incorporating ${H}_{\mathrm{SO}}$ requires several changes in the MR-ADC implementation:
\begin{enumerate}
\item
${H}_{\mathrm{SO}}$ modifies the amplitudes of correlation operator $T^{(k)}$ (\cref{eq:amp_form}) by entering the amplitude equations \eqref{eq:amp_eq} for the single and semi-internal double excitations.
Following the standard NEVPT2 notation,\cite{Angeli:2001p10252} these amplitudes belong to the $[\pm1']$ and $[0']$ excitation classes and can be denoted as $t^{a(k)}_i$, $t^{x(k)}_i$, $t^{a(k)}_x$, $t^{ay(k)}_{ix}$, $t^{yz(k)}_{ix}$, and $t^{az(k)}_{xy}$ using the orbital index labels in \cref{fig:orbital_diagram}. 
Due to the SOMF approximation, the amplitude equations for other classes ($[0]$, $[\pm1]$, $[\pm2]$) remain unaffected. 
These amplitudes describe double excitations involving at least two non-active molecular orbitals that cannot couple with the single-excitation spin--orbit operator ${H}_{\mathrm{SO}}$ (\cref{eq:h_2c_so_general}) when it enters \cref{eq:amp_form}.
As in our nonrelativistic implementation,\cite{Chatterjee2019:p59085924, Chatterjee2020:p63436357} the second-order correlation operator $T^{(2)}$ has negligible contributions to the MR-ADC matrices up to the MR-ADC(2)-X level of theory.
For this reason, we include only one class of second-order correlation amplitudes ($t^{a(2)}_i$) to ensure consistency with the single-reference ADC approximations.

\item
Since ${H}_{\mathrm{SO}}$ contains terms with all active indices, a new class of internal single excitations ($t^{y(1)}_x$, $x > y$) is introduced. 
These correlation amplitudes are necessary to account for the active-space spin--orbit coupling effects in the reference wavefunction and to ensure that the effective Hamiltonian matrix $\mathbf{M}_{\pm}$ is complex-Hermitian.
For additional details and derivation of $t^{y(1)}_x$ amplitude equations, we refer the readers to the Appendix. 

\item
Finally, the spin--orbit contributions to $T^{(k)}$ and $V$ modify the $\mathbf{M}_{\pm}$  and $\mathbf{T}_{\pm}$ matrix elements.
Implementation of these new contributions requires properly treating complex conjugation and permutational symmetry of complex-valued tensors.
\end{enumerate}

\begin{table*}[t]
	\label{tab:method_table}
	\caption{Spin--orbit SR- and MR-ADC methods implemented in this work.
		X2C-1e stands for the spin-free (SF) exact two-component approach of Liu and co-workers.\cite{Li:2012p154114}
		For the discussion of spin--orbit (SO) Hamiltonians and other details see \cref{sec:theory:soc_mradc}.
	}
	\label{tab:methods}
	\setstretch{1}
	\small
	\centering
	\begin{threeparttable}
		\begin{tabular}{lcc}
			\hline\hline
			Method 			& SF Hamiltonian		 & SO Hamiltonian\\ 
			\hline
			BP-(EA/IP)-(SR/MR)ADC(2) 		& X2C-1e 			& BP 			\\
			BP-(EA/IP)-(SR/MR)ADC(2)-X 		& X2C-1e 			& BP 			\\
			DKH1-(EA/IP)-(SR/MR)ADC(2)		& X2C-1e 			& DKH1		\\
			DKH1-(EA/IP)-(SR/MR)ADC(2)-X 	& X2C-1e 			& DKH1 		\\
			DKH2-(EA/IP)-(SR/MR)ADC(2) 		& X2C-1e + DKH2 	& DKH2 		\\
			DKH2-(EA/IP)-(SR/MR)ADC(2)-X 	& X2C-1e + DKH2 	& DKH2 		\\	
			\hline\hline \\
		\end{tabular}
	\end{threeparttable}
\end{table*}

\cref{tab:methods} summarizes the capabilities of our spin--orbit MR-ADC implementation, which allows to calculate electron-attached (EA) and ionized (IP) states using three variants of relativistic Hamiltonians (BP/DKH1/DKH2) up to the MR-ADC(2)-X level of theory.
Our implementation supports both CASSCF and restricted Hartree--Fock (RHF) reference wavefunctions and can be used to perform spin--orbit SR-ADC calculations for molecules with a closed-shell reference state. 
Although the MR-ADC($n$) methods developed in this work are perturbative in nature, they deliver the exact energies of SOMF BP/DKH1/DKH2 Hamiltonian when all orbitals are included in the active space starting with the first-order approximation ($n\ge 1$). 
Our current implementation is restricted to non-degenerate reference states due to the state-specific nature of correlation amplitudes determined from \cref{eq:amp_eq}.
A generalization of this approach to degenerate reference states will be reported in a forthcoming publication.

In the following sections, we present a benchmark study of the relativistic ADC methods, starting with a brief summary of computational details. 

\section{Computational details}
\label{sec:comp_details}
The spin--orbit  EA/IP-ADC methods were implemented in the development version of \textsc{Prism}.\cite{Sokolov:2023prism} 
All one- and two-electron integrals and the CASSCF reference wavefunctions were computed using \textsc{Pyscf}.\cite{Sun:2020p024109} 
The matrix elements of DKH1 Hamiltonian were computed by interfacing \textsc{Prism} with \textsc{Socutils}.\cite{Wang:2022socutils,Wang:2023p848855} 
The DKH2 matrix elements were implemented in a local version of \textsc{Socutils}.\cite{Majumder2024:p46764688} 

We performed four sets of benchmark calculations.
In \cref{sec:results_1}, we assess the accuracy of spin--orbit EA/IP-ADC methods for predicting zero-field splitting in the ${}^2P$ and ${}^2\Pi$ states of main group atoms and diatomics.
Next, in \cref{sec:results_2}, we carry out benchmark calculations for the transition metal atoms with $d^1$ and $d^9$ electronic configurations.
In \cref{sec:results_3}, we simulate the photoelectron spectra of cadmium halides (\ce{CdX2}, X = Cl, Br, I) using the IP-ADC methods. 
Finally, in \cref{sec:results_4}, we compute the photoelectron spectra of methyl iodide (\ce{CH3I}) at equilibrium and along the C--I bond dissociation.

All electrons were correlated in all ADC calculations.  
For an open-shell system containing $N$ electrons, the EA/IP-ADC results were computed starting with the $(N\mp1)$-electron lowest-energy singlet reference state.
The geometries, active spaces, and CASCI states ($\ket{\Psi_{\pm I}}$ in \cref{fig:excitations}) chosen for each calculation are provided in the Supplementary Information.
The MR-ADC calculations were performed using the $\eta_s=10^{-5}$ and  $\eta_d=10^{-10}$ parameters to remove linearly dependent semiinternal and double excitations, respectively.\cite{Chatterjee2019:p59085924,Chatterjee2020:p63436357}

For the main group elements and diatomics (\cref{sec:results_1}), we utilized the ANO-RCC-VTZP basis set.\cite{Roos:2004p28512858} 
The diatomic bond lengths were set to their experimental values,\cite{Huber:1982p298} which are provided in the Supplementary Information. 
The calculations of transition metal atoms with the $d^1$ and $d^9$ electronic configurations (\cref{sec:results_2}) were performed using the all electron X2C-TZVPall-2c basis set.\cite{Pollak:2017p3696}
To compute the photoelectron spectra of cadmium halides (\cref{sec:results_3}), we employed the X2C-QZVPall basis set\cite{Franzke2020:p56585674} and structural parameters from Ref.\@ \citenum{Scherpelz2016:p35233544}. 
The \ce{CdX2} experimental photoelectron spectra were digitized using the WebPlotDigitizer\cite{WebPlotDigitizer} from the data reported in Refs.\@ \citenum{Bristow1983:p263275} and \citenum{Kettunen2011:p901907}.

Finally, for the simulations of \ce{CH3I} photoelectron spectra (\cref{sec:results_4}) we used the X2C-TZVPall basis set.\cite{Pollak:2017p3696}
The \ce{CH3I} equilibrium geometry was optimized using density functional theory with the B3LYP functional\cite{Stephens1994:p1162311627} and the def2-TZVP basis set.\cite{Peterson2003:p1111311123,Weigend2005:p32973305} 
The reference CASSCF wavefunctions were calculated for the lowest-energy singlet state incorporating 6 electrons in 7 active orbitals (6e, 7o), which included the lone pairs of the iodine atom, the $\sigma$-bonding and antibonding C--I orbitals, and three more antibonding orbitals localized on the \ce{CH3} group.
Photoelectron spectra were simulated for the equilibrium, stretched, and completely dissociated \ce{CH3I} structures.
In the stretched geometry, the C--I bond was elongated by a factor of two relative to its equilibrium value ($r_e$), keeping the structure of \ce{CH3} group frozen (pyramidal). 
For the dissociated geometry (\ce{CH3}+I), the C--I distance was set to $\sim$6.7 \AA\ and the \ce{CH3} fragment was fully optimized at the CCSD(T)/def2-TZVP level of theory in a separate calculation without the I atom being present.
These geometries are reported in the Supplementary Information.

\section{Results and Discussion}
\label{sec:results}

\subsection{Zero-field splitting in main group atoms and diatomics}
\label{sec:results_1}

\begin{table*}[t!]
	\caption{
		Zero-field splitting (\cm) in the  ${}^2P$ states of main group atoms and the ${}^2\Pi$ states of diatomics computed using the spin--orbit EA-MR-ADC methods with the BP, DKH1, or DKH2 spin--orbit Hamiltonians. 
		All calculations employed the uncontracted ANO-RCC-VTZP basis set. 
	}
	\label{tab:eamradc}
	\setstretch{1}
	\scriptsize
	\centering
	\begin{threeparttable}
		\begin{tabular}{lccccccc}
			\hline\hline
			System			& BP-EA-		& DKH1-EA-		& DKH2-EA-		& BP-EA-			& DKH1-EA-		& DKH2-EA-		& Experiment\tnote{a}	\\ 
			& MR-ADC(2)	& MR-ADC(2)	& MR-ADC(2)	& MR-ADC(2)-X		& MR-ADC(2)-X		& MR-ADC(2)-X 		&			\\			
			\hline
			\ce{B} 			& 13.2		& 13.2		& 13.2      	& 13.8			& 13.8			& 13.8			& 15.0		\\	
			\ce{Al}			& 112		& 112 		& 112		& 117			& 116			& 116			& 112		\\
			\ce{Ga}		   	& 1045 		& 942 		& 949		& 998			& 899			& 906			& 826		\\
			\ce{In} 			& \tnote{b}			& 2796 		& 2843		& \tnote{b}				& 2756				& 2802			& 2213		\\
			\ce{Na}			& 13.0 		& 12.9 		& 12.9 		& 13.9			& 13.8			& 13.8			& 17.2		\\
			\ce{K}			& 43 			& 43 			& 43 			& 55				& 55 			& 55 			& 58			\\
			\ce{Rb}			& \tnote{b}			& 237  			& 239			& \tnote{b}				& 264			& 267			& 238		\\
			\ce{Cs}			& \tnote{b}			& 474			& 474			& \tnote{b}				& 584			& 595			& 554		\\
			\ce{CH}			& 26 			& 26 			& 26			& 26				& 26				& 26				& 28			\\
			\ce{SiH} 			& 135 		& 134 		& 134 		& 138			& 137			& 137			& 143 		\\
			\ce{GeH} 			& 1118 		& 894 		& 901 		& 1120       			& 892			& 900			& 893 		\\
			\ce{SnH} 			& \tnote{b}			& 2304		& 2344			& \tnote{b}			& 2361				& 2402				& 2178 		\\
			\ce{BeH}			& 1.76		& 1.76 		& 1.76 		& 1.88			& 1.88			& 1.88			& 2.14 		\\
			\ce{MgH} 			& 33 			& 33 			& 33  			& 36				&  36				& 36				& 35			\\
			\ce{CaH} 			& 77 			& 76 			& 76 			& 82				& 80				& 81 				& 79			\\
			\ce{SrH}			& \tnote{b} 			& 296 			& 261	 		& \tnote{b}				& 259			& 261			& 300 		\\
			\hline\hline
		\end{tabular}
		\begin{tablenotes}
			\item[a] Experimental results are from Refs. \citenum{huber:2013p298} and \citenum{NIST_ASD}.
			\item[b] Unphysical results encountered when using the BP Hamiltonian.
		\end{tablenotes}
	\end{threeparttable}
\end{table*}

\begin{table*}[t!]
	\caption{
		Zero-field splitting (\cm) in the  ${}^2P$ states of main group atoms and the ${}^2\Pi$ states of diatomics computed using the spin--orbit EA-SR-ADC methods with the BP, DKH1, or DKH2 spin--orbit Hamiltonians. 
		All calculations employed the uncontracted ANO-RCC-VTZP basis set. 
	}
	\label{tab:easradc}
	\setstretch{1}
	\scriptsize
	\centering
	\begin{threeparttable}
		\begin{tabular}{lccccccc}
			\hline\hline
			System			& BP-EA-		& DKH1-EA-		& DKH2-EA-		& BP-EA-			& DKH1-EA-		& DKH2-EA-		& Experiment\tnote{a}	\\ 
			& SR-ADC(2)	& SR-ADC(2)	& SR-ADC(2)	& SR-ADC(2)-X		& SR-ADC(2)-X		& SR-ADC(2)-X 		&			\\			
			\hline
			\ce{B} 			& 14.0		& 14.0		& 14.3	      	& 15.5				& 16.0				& 15.4				& 15.0 			\\	
			\ce{Al}			& 111		& 109 		& 103		& 109			& 115			& 109			& 112		\\
			\ce{Ga}		   	& 937 		& 845 		& 852		& 981			& 884			& 892			& 826		\\
			\ce{In} 			& \tnote{b}			& 2416 		& 2456		& \tnote{b}				& 2518			& 2559			& 2213		\\
			\ce{Na}			& 15.5 		& 15.5 		& 15.5 		& 16.1			& 16.0			& 16.0			& 17.2	\\
			\ce{K}			& 58 			& 57 			& 57 			& 61				& 60 				& 60 				& 58			\\
			\ce{Rb}			& \tnote{b}			& 238			& 240			& \tnote{b}				& 248				& 251				& 238		\\
			\ce{Cs}			& \tnote{b}			& 585			& 597			& \tnote{b}				& 610				& 623				& 554		\\
			\ce{CH}			& 26 			& 26 			& 26			& 29				& 29				& 29				& 28			\\
			\ce{SiH} 			& 142 		& 140 		& 140 		& 150			& 149			& 149			& 143 		\\
			\ce{GeH} 			& 1151 		& 919 		& 927 		& 1214      			& 969			& 978			& 893 		\\
			\ce{SnH} 			& \tnote{b}			& 2412		& 2454		& \tnote{b}				& 2534			& 2578			& 2178 		\\
			\ce{BeH}			& 1.74 		& 1.74 		& 1.75 		& 1.85			& 1.84			& 1.85			& 2.14 		\\
			\ce{MgH} 			& 34 			& 32 			& 33  		& 34				& 34 				& 34				& 35			\\
			\ce{CaH} 			& 83 			& 81			& 81 			& 86				& 85				& 85 				& 79			\\
			\ce{SrH}			& \tnote{b} 			& 291 		& 294 		& \tnote{b}				& 308			& 311			& 300 		\\
			\hline\hline
		\end{tabular}
		\begin{tablenotes}
			\item[a] Experimental results are from Refs. \citenum{huber:2013p298} and \citenum{NIST_ASD}.
			\item[b] Unphysical results encountered when using the BP Hamiltonian.
		\end{tablenotes}
	\end{threeparttable}
\end{table*}

\begin{table*}[t!]
	\caption{
		Zero-field splitting (\cm) in the  ${}^2P$ states of main group atoms and the ${}^2\Pi$ states of diatomics computed using the spin--orbit IP-MR-ADC methods with the BP, DKH1, or DKH2 spin--orbit Hamiltonians. 
		All calculations employed the uncontracted ANO-RCC-VTZP basis set. 
	}
	\label{tab:ipmradc}
	\setstretch{1}
	\scriptsize
	\centering
	\begin{threeparttable}
		\begin{tabular}{lccccccc}
			\hline\hline
			System			& BP-IP-		& DKH1-IP-		& DKH2-IP-		& BP-IP-			& DKH1-IP-		& DKH2-IP-		& Experiment\tnote{a}	\\ 
			& MR-ADC(2)	& MR-ADC(2)	& MR-ADC(2)	& MR-ADC(2)-X		& MR-ADC(2)-X		& MR-ADC(2)-X 		&			\\			
			\hline
			\ce{F} 			& 385		& 384		& 384	      	& 389			& 388			& 388			& 404		\\	
			\ce{Cl}			& 885		& 873 		& 875		& 892			& 880			& 880			& 882		\\
			\ce{Br}		   	& 4014		& 3672 		& 3708		& \tnote{b} 				& 3701			& 3737			& 3685		\\
			\ce{I} 			& 9980		& 7533 		& 7661		& \tnote{b}				& 7593			& 7722			& 7603		\\
			\ce{Ne+}			& 757 		& 754 		& 755 		& 760			& 757			& 758			& 780		\\
			\ce{Ar+}			& 1417 	& 1395 		& 1256		& 1425			& 1403			& 1410			& 1432		\\
			\ce{Kr+}			& 5906		& 5369		& 5423		& \tnote{b} 				& 5386			& 5441			& 5370		\\
			\ce{Xe+}			& 12780	& 9832			& 9996			& \tnote{b}			& 9751				& 9914			& 10537		\\
			\ce{Rn+}			& \tnote{b}			& 27208		& 27654		& \tnote{b}				& 27388			& 27842			& 30895		\\
			\ce{OH}			& 128	& 128 		& 128		& 130			& 130			& 130			& 139		\\
			\ce{SH} 			& 348 		& 344 		& 345 		& 341			& 337			& 338			& 377 		\\
			\ce{SeH} 			& 1754 		& 1623 		& 1640 		& \tnote{b}       			& 1571			& 1585			& 1764 		\\
			\ce{TeH} 			& 4294		& 3438		& 3492		& \tnote{b}				& 3335			& 3386			& 3816 		\\
			\ce{HF+}			& 279 		& 278 		& 279 		& 281			& 280			& 281			& 293 		\\
			\ce{HCl+} 			& 613		& 605 		& 607  		& 616			& 608 			& 610			& 648		\\
			\ce{HBr+} 			& 2717		& 2502		& 2525		& \tnote{b} 				& 2490			& 2513			& 2653		\\
			\ce{HI+}			& 6179		& 4915 		& 4990 		& \tnote{b}				& 4866			& 4941			& 5400 		\\
			\hline\hline
		\end{tabular}
		\begin{tablenotes}
			\item[a] Experimental results are from Refs. \citenum{huber:2013p298} and \citenum{NIST_ASD}.
			\item[b] Unphysical results encountered when using the BP Hamiltonian.
		\end{tablenotes}
	\end{threeparttable}
\end{table*}

\begin{table*}[t!]
	\caption{
		Zero-field splitting (\cm) in the  ${}^2P$ states of main group atoms and the ${}^2\Pi$ states of diatomics computed using the spin--orbit IP-SR-ADC methods with the BP, DKH1, or DKH2 spin--orbit Hamiltonians. 
		All calculations employed the uncontracted ANO-RCC-VTZP basis set. 
	}
	\label{tab:ipsradc}
	\setstretch{1}
	\scriptsize
	\centering
	\begin{threeparttable}
		\begin{tabular}{lccccccc}
			\hline\hline
			System			& BP-IP-		& DKH1-IP-	& DKH2-IP-	& BP-IP-			& DKH1-IP-		& DKH2-IP-		& Experiment\tnote{a}	\\ 
			& SR-ADC(2)	& SR-ADC(2)	& SR-ADC(2)	& SR-ADC(2)-X		& SR-ADC(2)-X		& SR-ADC(2)-X 		&			\\			
			\hline
			\ce{F} 			& 382		& 389		& 381	      	& 439			& 437			& 438			& 404		\\	
			\ce{Cl}			& 849		& 837 		& 840		& 917			& 904			& 907			& 882		\\
			\ce{Br}		   	& 3783		& 3478 		& 3511		& \tnote{b} 				& 3703			& 3738			& 3685		\\
			\ce{I} 			& 8926		& 6997 		& 7115		& \tnote{b}				& 7383			& 7504			& 7603		\\
			\ce{Ne+}			& 761 		& 760 		& 760 		& 815			& 812			& 813			& 780		\\
			\ce{Ar+}			& 1419 		& 1397 		& 1401		& 1473			& 1451			& 1455			& 1432		\\
			\ce{Kr+}			& 5703		& 5208		& 5261		& \tnote{b} 				& 5364			& 5417			& 5370		\\
			\ce{Xe+}			& 12957 		& 9988		& 10156		& \tnote{b}				& 10232			& 10404			& 10537		\\
			\ce{Rn+}			& 25547		& 25546		& 25949		& \tnote{b}				& 26287			& 26729			& 30895		\\
			\ce{OH}			& 132		& 132 		& 132		& 153			& 152			& 152			& 139		\\
			\ce{SH} 			& 359 		& 355 		& 356 		& 388			& 382			& 383			& 377 		\\
			\ce{SeH} 			& 1774 		& 1640 		& 1658 		& \tnote{b}       			& 1746			& 1761			& 1764 		\\
			\ce{TeH} 			& 4269		& 3432		& 3485		& \tnote{b}				& 3605			& 3660			& 3816 		\\
			\ce{HF+}			& 283 		& 283 		& 283 		& 312			& 311			& 311			& 293 		\\
			\ce{HCl+} 			& 635		& 626 		& 628  		& 662			& 654 			& 655			& 648		\\
			\ce{HBr+} 			& 2771		& 2553		& 2578		& \tnote{b} 				& 2634			& 2659			& 2653		\\
			\ce{HI+}			& 6261		& 4998 		& 5074 		& \tnote{b}				& 5118			& 5196			& 5400 		\\
			\hline\hline
		\end{tabular}
		\begin{tablenotes}
			\item[a] Experimental results are from Refs. \citenum{huber:2013p298} and \citenum{NIST_ASD}.
			\item[b] Unphysical results encountered when using the BP Hamiltonian.
		\end{tablenotes}
	\end{threeparttable}
\end{table*}

\begin{figure*}[t!]
	\includegraphics[width=0.8\textwidth]{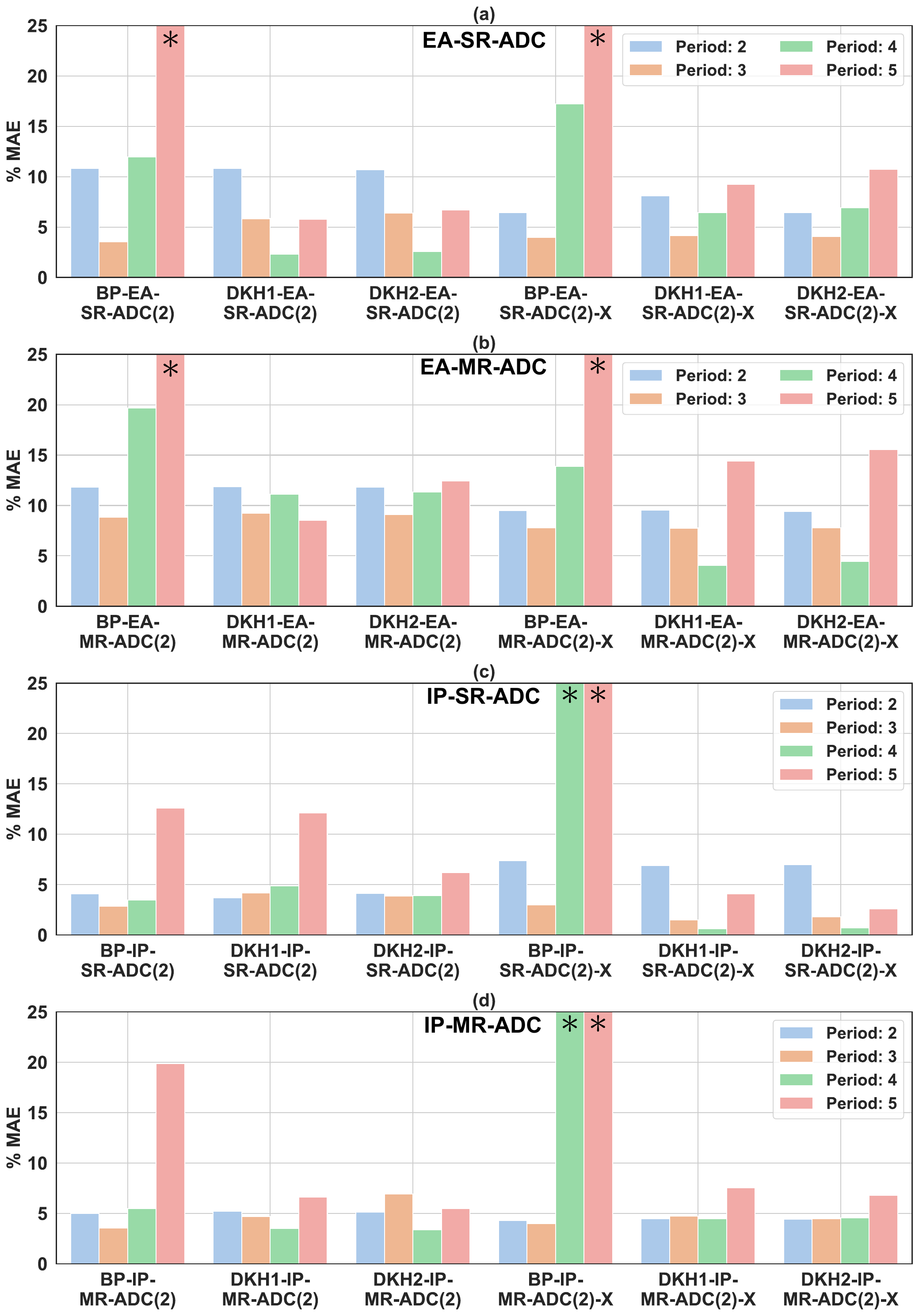} 
	\captionsetup{justification=justified,singlelinecheck=false,font=footnotesize}
	\caption{
		Percent mean absolute errors (\% MAE) in the zero-field splitting of main group atoms and diatomics calculated using the spin--orbit EA/IP-ADC methods for the different rows of periodic table relative to the experimental measurements.
		Bars that exceed the scale are indicated with asterisks. 
		See \cref{tab:eamradc,tab:easradc,tab:ipsradc,tab:ipmradc} for the data on individual systems.
	} 
	\label{fig:all_errors}
\end{figure*}

\begin{table*}[t]
	\caption{
	Mean absolute errors (\cm) in the zero-field splitting of main group atoms and diatomics calculated using the spin--orbit EA/IP-ADC methods for periods 2 to 5 of the periodic table relative to the experimental measurements.
	See \cref{tab:eamradc,tab:easradc,tab:ipsradc,tab:ipmradc} for the data on individual systems.
	}
	\label{tab:all_errors_cm}
	\setstretch{1}
	\footnotesize
	\centering
	\begin{threeparttable}
		\begin{tabular}{lcccc}
			\hline\hline
			Method & Period 2 &	Period 3 & Period 4 & Period 5	\\			
			\hline
			BP-EA-SR-ADC(2)           &1.1 & 1.2 & 93.4 & \tnote{a} \\			
			BP-EA-SR-ADC(2)-X       & 0.5 & 2.9 & 121.5 & \tnote{a} \\
			BP-EA-MR-ADC(2)           & 1.2 & 3.6 & 114.9 & \tnote{a} \\
			BP-EA-MR-ADC(2)-X       & 1.2 & 3.6 & 101.2 & \tnote{a} \\			
			DKH1-EA-SR-ADC(2)       & 1.1 & 2.7 & 12.0 & 111.7  \\			
			DKH1-EA-SR-ADC(2)-X  & 0.8 & 2.8 & 35.5 &169.8 \\
			DKH1-EA-MR-ADC(2)      & 1.3 & 3.9 & 33.7 & 178.8 \\			
			DKH1-EA-MR-ADC(2)-X  & 1.3 & 3.7 & 19.6 & 198.4 \\			
			DKH2-EA-SR-ADC(2)      & 1.2 & 3.8 & 15.7 & 132.0 \\			
			DKH2-EA-SR-ADC(2)-X  & 0.5 & 2.8 & 39.4 &192.3 \\			
			DKH2-EA-MR-ADC(2)      & 1.2 & 3.8 & 37.4 & 209.0 \\			
			DKH2-EA-MR-ADC(2)-X  & 1.2 & 3.7 & 22.9 & 220.2 \\			
			BP-IP-SR-ADC(2)             & 14.5 & 19.3 & 139.8 & 1264.3 \\			
			BP-IP-SR-ADC(2)-X         & 25.8 &25.3 & \tnote{a} & \tnote{a} \\			
			BP-IP-MR-ADC(2)            & 16.6 & 20.1 & 234.7 & 1469.3 \\			
			BP-IP-MR-ADC(2)-X        & 14.2 & 21.3 & \tnote{a} & \tnote{a} \\			
			DKH1-IP-SR-ADC(2)        & 13.0 & 31.0 & 148.3 & 485.3 \\			
			DKH1-IP-SR-ADC(2)-X    & 24.0 & 13.0 & 15.3 & 254.5 \\			
			DKH1-IP-MR-ADC(2)       & 17.7 & 30.2 & 76.5 & 409.5 \\			
			DKH1-IP-MR-ADC(2)-X   & 15.2 & 27.6 & 97.2 & 452.8 \\			
			DKH2-IP-SR-ADC(2)        & 14.9 & 28.3 & 116.1 & 381.4 \\			
			DKH2-IP-SR-ADC(2)-X    & 24.5 & 15.3 & 27.3 & 148.1 \\			
			DKH2-IP-MR-ADC(2)       & 17.3 & 63.8 & 82.1 & 333.1 \\			
			DKH2-IP-MR-ADC(2)-X   & 14.9 & 25.0 & 110.6 & 407.6 \\			
			\hline\hline
		\end{tabular}
		\begin{tablenotes}
		\item[a] Unphysical results encountered when using the BP Hamiltonian.
		\end{tablenotes}
	\end{threeparttable}
\end{table*}
	
We begin with a benchmark of spin--orbit ADC approximations for calculating the zero-field splitting (ZFS) in main group atoms and diatomic molecules that do not exhibit multireference effects. 
\cref{tab:eamradc,tab:easradc} compare the results of EA-ADC methods with available experimental data\cite{Cao:2017p3713,NIST_ASD} for the group 1 and 13 atoms and group 2 and 14 hydrides.
The IP-ADC benchmark calculations (\cref{tab:ipmradc,tab:ipsradc}) were performed for the group 17 atoms, group 18 cations, as well as group 16 neutral and group 17 cationic hydrides.
For an atom or molecule with $N$ electrons, the EA/IP-ADC calculations were performed for the lowest-energy term of ${}^2P$ or ${}^2\Pi$ symmetry starting with the ($N\mp 1$) singlet reference wavefunction. 
Additional computational details can be found in \cref{sec:comp_details} and the Supporting Information. 

The benchmark results are summarized in \cref{fig:all_errors} and \cref{tab:all_errors_cm} where the EA/IP-ADC mean absolute errors (\mae) in \% and \cm are calculated relative to the experimental data for each row of periodic table. 
For the second- and third-period elements, the computed ZFS show little dependence on the choice of two-component spin--orbit Hamiltonian (BP, DKH1, and DKH2).
Starting with the fourth period, the BP-ADC methods deteriorate in accuracy and tend to produce unphysical results with large negative excitation energies, likely due to the lack of variational lower bound of the BP Hamiltonian (\cref{sec:theory:soc_mradc}).
The DKH1- and DKH2-ADC calculations do not exhibit these issues and are significantly more accurate compared to BP-ADC for heavier elements.

To compare the accuracy of ADC levels of theory in predicting the ZFS of main group elements, we focus on the DKH2 results in \cref{fig:all_errors} and \cref{tab:all_errors_cm}. 
For periods 2 and 3, all DKH2-EA-ADC methods show similar accuracy with \mae of $\sim$ 1 and 3 \cm, which represents $\sim$ 5 to 10 \% error relative to experimental ZFS due to weak spin--orbit coupling in these systems.
In periods 4 and 5, the DKH2-EA-ADC \mae range from 16 to 39 \cm (2.6 to 11.3 \%) and from 132 to 220 \cm (5.8 to 15.5 \%), respectively.
Since the molecules in this benchmark set do not exhibit multireference effects, the EA-SR-ADC methods are competitive in accuracy to EA-MR-ADC, often showing better performance.  
The DKH2-EA-SR-ADC(2) method has the smallest \mae for periods 4 and 5, despite being the lowest level of theory out of four DKH2-EA-ADC approximations. 

The DKH2-IP-ADC methods show somewhat larger errors in ZFS compared to DKH2-EA-ADC, which represent a smaller \% fraction ($\sim$ 2 to 6 \%)  of the experimental reference data. 
Going down the periodic table, the DKH2-IP-ADC \mae ranges are 14.9 -- 24.5, 15.3 -- 63.8, 27.3 -- 116.1, and 148.1 -- 407.6 \cm for periods 2, 3, 4, and 5, respectively (\cref{tab:all_errors_cm}).
The DKH2-IP-SR-ADC(2)-X and DKH2-IP-MR-ADC(2) methods tend to show smaller \mae for periods 4 and 5 within a limited scope of our benchmark study.

\subsection{Spin--orbit coupling in $d^1$ and $d^9$ transition metal atoms}
\label{sec:results_2}

\begin{table*}[t]
	\caption{
		Zero-field splitting (\cm) in the ground ${}^2D$ term of \ce{Sc}, \ce{Y} and \ce{La} atoms computed using the spin--orbit EA-ADC methods.
		For comparison, we also include data from DKH2-QDNEVPT2\cite{Majumder2024:p46764688} and X2C-MRCISD.\cite{Hu:2020p29752984}
		All calculations employed the X2C-TZVPall-2c basis set. 
		Shown in parentheses are the \% errors with respect to experimental results.\cite{Sugar:1985,Koseki:2019p23252339,NIST_ASD}
	}
	\label{tab:ea_mradc_metal}
	\setstretch{1}
	\footnotesize
	\centering
	\begin{threeparttable}
		\begin{tabular}{lccc}
			\hline\hline  			
			Method				& \ce{Sc}				& \ce{Y}				& \ce{La}				\\			
			\hline
			DKH2-QDNEVPT2\cite{Majumder2024:p46764688}  & 141 (16.3)			  &  428 (19.2)			  & 897 (14.9)				\\
			X2C-MRCISD\cite{Hu:2020p29752984}   		& 186 (10.2)			&  524 (1.1)			  & 936 (11.2)				\\
			BP-EA-SR-ADC(2)		 		& 108 (35.5)			& 381 (28.2)			& \tnote{a}			\\
			DKH1-EA-SR-ADC(2)		 & 110 (34.7)			& 392 (26.1)			& 987 (6.2)				\\
			DKH2-EA-SR-ADC(2) 		& 110 (34.8)			& 391 (26.3)			& 987 (6.2)				\\
			BP-EA-SR-ADC(2)-X   	 & 131 (22.0)			   & 450 (15.1)			  & \tnote{a}				\\
			DKH1-EA-SR-ADC(2)-X   & 132 (21.3)			   & 457 (13.7)			  &1094 (3.9)				\\
			DKH2-EA-SR-ADC(2)-X   & 132 (21.4)			   & 457 (13.8)			  & 1095 (4.0)				\\
			BP-EA-MR-ADC(2) 		  & 109 (35.3)			  & 385 (27.4)			& 973 (7.6)				\\
			DKH1-EA-MR-ADC(2) 		& 110 (34.6)			& 396 (25.3)			& 1002 (4.9)				\\
			DKH2-EA-MR-ADC(2) 		& 110 (34.6)			& 395 (25.5)		   & 1002(4.9)				\\
			BP-EA-MR-ADC(2)-X         & 132 (21.3)			 & 454 (14.3)			  & \tnote{a}				\\
			DKH1-EA-MR-ADC(2)-X    & 133 (20.7)			   & 461 (12.9)			  & 1089 (3.4)				\\
			DKH2-EA-MR-ADC(2)-X   & 133 (20.7)			  &  461 (13.0)			  & 1090 (3.5)				\\
			Experiment	 						&  168      	    					 & 530 				  &  1053					\\
			\hline\hline
		\end{tabular}
			\begin{tablenotes}
				\item[a] Unphysical results encountered when using the BP Hamiltonian.
			\end{tablenotes}
	\end{threeparttable}
\end{table*}

\begin{table*}[t]
	\caption{
		Zero-field splitting (\cm) in the excited ${}^2D$ term of \ce{Cu}, \ce{Ag}, and \ce{Au} atoms computed using the spin--orbit IP-ADC methods.
		All calculations employed the X2C-TZVPall-2c basis set. 
		Shown in parentheses are the \% errors with respect to experimental results.\cite{Sugar1990:p527616,Pickering2001:p181185,Sansonetti2005:p15592259}
	}
	\label{tab:d9_adc}
	\setstretch{1}
	\footnotesize
	\centering
	\begin{threeparttable}
		\begin{tabular}{lccc}
			\hline\hline
			Method				& \ce{Cu}				& \ce{Ag}				& \ce{Au}				\\			
			\hline
			BP-IP-SR-ADC(2)			&1787 (12.5)			&4071 (9.0)			&11547 (5.9)				\\
			DKH1-IP-SR-ADC(2)		&1785 (12.6)			&4027 (9.9)			&11105 (9.5)				\\
			DKH2-IP-SR-ADC(2) 		&1786 (12.6)			&4034 (9.8)			&11168 (9.0)				\\
			BP-IP-SR-ADC(2)-X 			&2181 (6.8)				&4727 (5.7)			&\tnote{a} 					\\
			DKH1-IP-SR-ADC(2)-X 		&2177 (6.6)				&4659 (4.2)			&13030 (6.2)				\\
			DKH2-IP-SR-ADC(2)-X 		&2178 (6.6)				&4668 (4.4)			&13108 (6.8)				\\
			BP-IP-MR-ADC(2)			&1927 (5.7)				&4291 (4.0)			&12109 (1.3)				\\
			DKH1-IP-MR-ADC(2) 		&1925 (5.8)				&4245 (5.1)			&11602 (5.5)				\\
			DKH2-IP-MR-ADC(2) 		&1926 (5.7)				&4251 (4.9)			&11666 (4.9)				\\
			BP-IP-MR-ADC(2)-X		&1984 (2.9)				&4344 (2.9)			&\tnote{a} 					\\
			DKH1-IP-MR-ADC(2)-X 		&2019 (1.1)				&4292 (4.0)			&11490 (6.4)				\\
			DKH2-IP-MR-ADC(2)-X 		&2021 (1.1)				&4299 (3.9)			&11547 (5.9)				\\
			Experiment	 				&2043		&4472			&12274		\\
			\hline\hline
		\end{tabular}
		\begin{tablenotes}
			\item[a] Unphysical results encountered when using the BP Hamiltonian.
		\end{tablenotes}
	\end{threeparttable}
\end{table*}

We now turn our attention to the transition metal atoms with the $d^1$ (ground-state Sc, Y, La) and $d^9$ (excited-state Cu, Ag, Au) electronic configurations. 
\cref{tab:ea_mradc_metal} reports the ZFS in the ground ${}^2D$ term of Sc, Y, and La atoms computed using the spin--orbit EA-ADC methods starting with the ${}^1S$ reference states of their cations. 
Earlier studies using two-component multireference configuration interaction (X2C-MRCISD)\cite{Hu:2020p29752984} and quasidegenerate N-electron valence perturbation theory (DKH2-QDNEVPT2)\cite{Majumder2024:p46764688} reported significant errors in the ZFS of these elements (\cref{tab:ea_mradc_metal}). 
For example, the variational X2C-MRCISD method shows the 10.2 and 11.2 \% errors in ZFS for Sc and La, respectively. 
The smallest error in the DKH2-QDNEVPT2 calculations is 14.9 \% (La).\cite{Majumder2024:p46764688}

For all $d^1$ atoms (Sc, Y, and La), the EA-SR-ADC and EA-MR-ADC methods show similar results at the same level of spin--orbit and dynamic correlation treatment.
The DKH-EA-ADC(2)-X family of methods exhibits the best performance predicting the ZFS of Sc, Y, and La within $\sim$ 21, 13, and 4 \% of the experimental data,\cite{Sugar:1985,Koseki:2019p23252339,NIST_ASD} respectively (\cref{tab:ea_mradc_metal}). 
For the La atom, the DKH-EA-ADC(2)-X methods outperform the X2C-MRCISD approach, likely due to a fortuitous error cancellation. 
When compared to DKH2-QDNEVPT2, DKH-EA-ADC(2)-X show better results for Y and La. 
The strict second-order approximations (DKH-EA-ADC(2)) exhibit significantly larger errors than their extended (-X) counterparts ($\sim$ 35, 26, and 5 \% for Sc, Y, and La).
As for the main group elements and diatomics (\cref{sec:results_1}), the BP spin--orbit Hamiltonian produces similar results to DKH1/DKH2 for lighter elements (Sc and Y) but is unreliable for the heavier La atom.

To assess the performance of spin--orbit IP-ADC approximations, we calculated the ZFS of Cu, Ag, and Au atoms in the excited ${}^2D$ term ($d^9$ electronic configuration) starting with the lowest-energy closed-shell anionic reference state (\cref{tab:d9_adc}).
In contrast to the  $d^1$ atoms, the IP-SR-ADC and IP-MR-ADC ZFS are significantly different, with the multireference approximations showing closer agreement with the experimental data.
The DKH-IP-MR-ADC(2)-X methods exhibit the best performance, predicting the ZFS of Cu, Ag, and Au within $\sim$ 1, 4, and 6 \% of their experimental values, respectively. 
DKH-IP-MR-ADC(2) yield similar results for Ag and Au but are somewhat less accurate for Cu with $\sim$ 6 \% error. 
The IP-SR-ADC results show much greater spread, changing significantly (by as much as 1940 \cm) from IP-SR-ADC(2) to IP-SR-ADC(2)-X.

Overall, our calculations highlight the importance of multireference effects  for simulating the ZFS in excited ${}^2D$ term of Cu, Ag, and Au.
These findings can be confirmed with the analysis of CASCI states in the MR-ADC calculations, which reveals that the ${}^2D$ excited states show significant mixing of electronic configurations with partially filled $d$-, $s$-, and $p$-shells that is particularly strong for Cu and Au. 
Consistent with this analysis, these two atoms display the largest difference in \% ZFS errors between the SR-ADC(2)-X and MR-ADC(2)-X methods. 

\subsection{Photoelectron spectra of cadmium halides}
\label{sec:results_3}

\begin{figure*}[t]
	\includegraphics[width=0.86\textwidth]{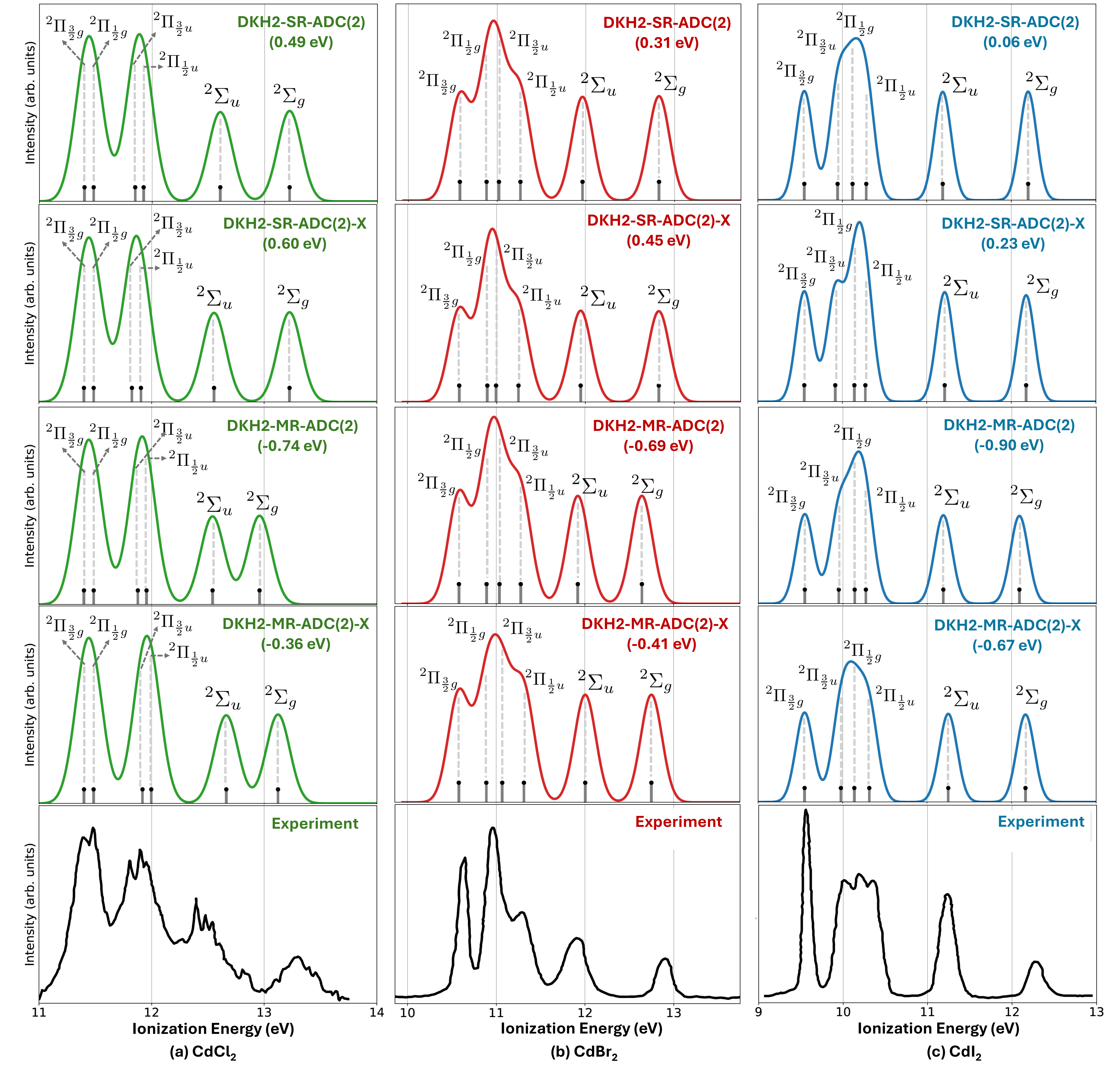} 
	\captionsetup{justification=justified,singlelinecheck=false,font=footnotesize}
	\caption{
	Photoelectron spectra of cadmium dihalides (\ce{CdX2}, \ce{X} = \ce{Cl}, \ce{Br}, and \ce{l}) simulated using IP-SR-ADC and IP-MR-ADC methods with the DKH2 spin--orbit Hamiltonian. 
	Calculated spectra were shifted to align them with the experimental spectra for the first peak.
	The shift value is indicated in parentheses for each spectrum. 
	Experimental spectra were digitized\cite{WebPlotDigitizer} and reprinted from Refs.\@ \citenum{Bristow1983:p263275} and \citenum{Kettunen2011:p901907} with permission from Elsevier and Wiley Materials. Copyright 1983 and 2011.
	} 
	\label{fig:cdx_mradc}
\end{figure*}

\begin{table*}[t]
	\caption{
		Relative energies ($\Delta E$, eV) of states in the \ce{CdX2+} molecules (X = Cl, Br, I) calculated using DKH2-IP-ADC methods in comparison to experimental data.\cite{Bristow1983:p263275,Kettunen2011:p901907} 
		For the ${}^2\Pi_{\frac{3}{2}g}$ state, vertical ionization energy (VIE, eV) is reported.
		All calculations employed the X2C-QZVPall basis set. 
	}
	\label{tab:cdx2_data}
	\setstretch{1}
	\footnotesize
	\centering
	\begin{threeparttable}
		\begin{tabular}{lcccccc}
			\hline\hline
			Molecule &Property &	SR-ADC(2)	& SR-ADC(2)-X	& MR-ADC(2)	& MR-ADC(2)-X	& Experiment	\\			
			\hline
			 \ce{CdCl2+}& VIE (${}^2\Pi_{\frac{3}{2}g}$)&	11.00	& 10.89	& 12.23	& 11.85	& 11.49	\\			
			 &$\Delta E({}^2\Pi_{\frac{1}{2}g} - {}^2\Pi_{\frac{3}{2}g})$ &	0.08	& 0.09	& 0.09	& 0.09	& $\lesssim$ 0.1	\\			
			&$\Delta E({}^2\Pi_{\frac{3}{2}u} - {}^2\Pi_{\frac{1}{2}g})$ &		0.37	& 0.34	&  0.39	&  0.43	& 0.40 	\\			
			&$\Delta E({}^2\Pi_{\frac{1}{2}u} - {}^2\Pi_{\frac{3}{2}u})$&	0.07	& 0.08	& 0.08	& 0.08	& $\lesssim$ 0.1	\\			
			&$\Delta E({}^2\Sigma_{u} - {}^2\Pi_{\frac{1}{2}u})$&		0.68	& 0.65	&  0.59	& 0.67	&  0.49	\\			
			&$\Delta E({}^2\Sigma_{g} - {}^2\Sigma_{u})$&		0.62	& 0.67	& 0.42	& 0.46	& 0.81	\\			
			 \ce{CdBr2+}&VIE (${}^2\Pi_{\frac{3}{2}g}$)&	10.27	& 10.12	& 11.28	& 10.99	& 10.58	\\			
			&$\Delta E({}^2\Pi_{\frac{1}{2}g}-{}^2\Pi_{\frac{3}{2}g})$&	0.30	& 0.32 & 0.31	& 0.31	& 0.31	\\			
			&$\Delta E({}^2\Pi_{\frac{3}{2}u}-{}^2\Pi_{\frac{1}{2}g})$ &	0.14	& 0.10	& 0.15	& 0.18	&  0.15	\\			
			&$\Delta E({}^2\Pi_{\frac{1}{2}u}-{}^2\Pi_{\frac{3}{2}u})$ &	0.24	& 0.25	& 0.24	& 0.24	& 0.21	\\			
			&$\Delta E({}^2\Sigma_{u}-{}^2\Pi_{\frac{1}{2}u})$ &	0.70	& 0.70	& 0.64	& 0.69	& 0.60	\\			
			&$\Delta E({}^2\Sigma_{g}-{}^2\Sigma_{u})$ &	0.86	& 0.88	& 0.72	& 0.75	& 1.01 	\\			
			 \ce{CdI2+}&VIE (${}^2\Pi_{\frac{3}{2}g}$)&	9.49	& 9.31	& 10.45	& 10.22	& 9.55 \\			
			 &$\Delta E({}^2\Pi_{\frac{3}{2}u}-{}^2\Pi_{\frac{3}{2}g})$ &	0.39	& 0.37	& 0.40	& 0.43	& 0.43	\\			
			 &$\Delta E({}^2\Pi_{\frac{1}{2}g}-{}^2\Pi_{\frac{3}{2}u})$ &	0.18	& 0.23	& 0.19	& 0.16	& 0.20 	\\			
			 &$\Delta E({}^2\Pi_{\frac{1}{2}u}-{}^2\Pi_{\frac{1}{2}g})$ &	0.16	& 0.13	& 0.13	& 0.18	& 0.17	\\			
			 &$\Delta E({}^2\Sigma_{u}-{}^2\Pi_{\frac{1}{2}u})$ &	0.90	& 0.94	& 0.92	& 0.93	& 0.86	\\			
			 &$\Delta E({}^2\Sigma_{g}-{}^2\Sigma_{u})$ &	1.01	& 0.96	& 0.90	& 0.91	& 1.05 	\\			
			\hline\hline
		\end{tabular}
	\end{threeparttable}
\end{table*}

In addition to charged excitation energies, the EA/IP-ADC methods provide straightforward access to transition probabilities that can be used to simulate photoelectron spectra. 
Here, we use our spin--orbit EA/IP-ADC implementation to compute the photoelectron spectra of linear cadmium halides (\ce{CdX2}, X = \ce{Cl}, \ce{Br}, \ce{I}).
Each molecule has a singlet ground state with the $(\sigma_g)^2(\sigma_u)^2(\pi_u)^4(\pi_g)^4$ electronic configuration in the order of increasing orbital energy.
Ionizing the doubly-degenerate $\pi_g$ and $\pi_u$ orbitals localized on the halogen atoms gives rise to four electronic states: ${}^2\Pi_{\frac{3}{2}g}$, ${}^2\Pi_{\frac{1}{2}g}$, ${}^2\Pi_{\frac{3}{2}u}$, and ${}^2\Pi_{\frac{1}{2}u}$.
The energy spacing and relative order of these states in \ce{CdX2+} depends on the strength of spin--orbit coupling that increases from X = \ce{Cl} to X = \ce{I}.

\cref{fig:cdx_mradc} compares the experimental photoelectron spectra\cite{Bristow1983:p263275,Kettunen2011:p901907} of \ce{CdX2} (X = \ce{Cl}, \ce{Br}, \ce{I}) with the results of DKH2-IP-MR/SR-ADC(2) and DKH2-IP-MR/SR-ADC(2)-X calculations. 
The simulated spectra were uniformly shifted to align their lowest-energy peak with the corresponding signal in the experimental data. 
Apart from the shift, all four levels of theory predict the same order of states and qualitatively reproduce the peak structure in experimental spectra.
For \ce{CdCl2}, four peaks are observed in the simulated and experimental photoelectron spectra.
The first two peaks correspond to two pairs of states (${}^2\Pi_{\frac{3}{2}g}$ -- ${}^2\Pi_{\frac{1}{2}g}$  and  ${}^2\Pi_{\frac{3}{2}u}$ -- ${}^2\Pi_{\frac{1}{2}u}$) with each pair split by $\lesssim$ 0.1 eV due to weak spin--orbit coupling.

Stronger zero-field splitting in \ce{CdBr2} and \ce{CdI2} merges the signals from ${}^2\Pi_{g}$  and ${}^2\Pi_{u}$ states into a broad band and reorders ${}^2\Pi_{\frac{1}{2}g}$ and ${}^2\Pi_{\frac{3}{2}u}$ in cadmium iodide.
The shape of this band in experimental spectra is qualitatively reproduced by all spin--orbit IP-ADC methods, suggesting that multireference effects are not important for the low-energy ionized states of cadmium halides. 
The IP-ADC calculations are also in a good agreement with the photoelectron spectra from spin--orbit self-consistent GW reported recently by Abraham et al.\cite{Abraham:2024p45794590}

\cref{tab:cdx2_data} reports the relative energies of \ce{CdX2+} (X = \ce{Cl}, \ce{Br}, \ce{I}) states in the experimental and simulated spectra. 
All spin--orbit ADC methods predict the relative spacing between the first four states within $\lesssim$ 0.06 eV of experimental measurements.
The ${}^2\Pi_{\frac{1}{2}u}$ -- ${}^2\Sigma_{u}$ energy separations are consistently overestimated in all ADC calculations by up to 0.2 eV.
The most significant deviations from experimental data are observed for the ${}^2\Sigma_{u}$ -- ${}^2\Sigma_{g}$ relative energies, which are systematically underestimated by 0.1 to 0.4 eV with errors increasing from X = I to X = Cl.
Due to the dissociative nature of  ${}^2\Sigma_{u}$ and ${}^2\Sigma_{g}$ states,\cite{Kettunen2011:p901907} accurately simulating their signals in photoelectron spectra may require considering the effects of nuclear dynamics, which are missing in our calculations.

\subsection{Photoelectron spectra of methyl iodide along bond dissociation}
\label{sec:results_4}

\begin{figure*}[t!]
	\includegraphics[width=0.9\textwidth]{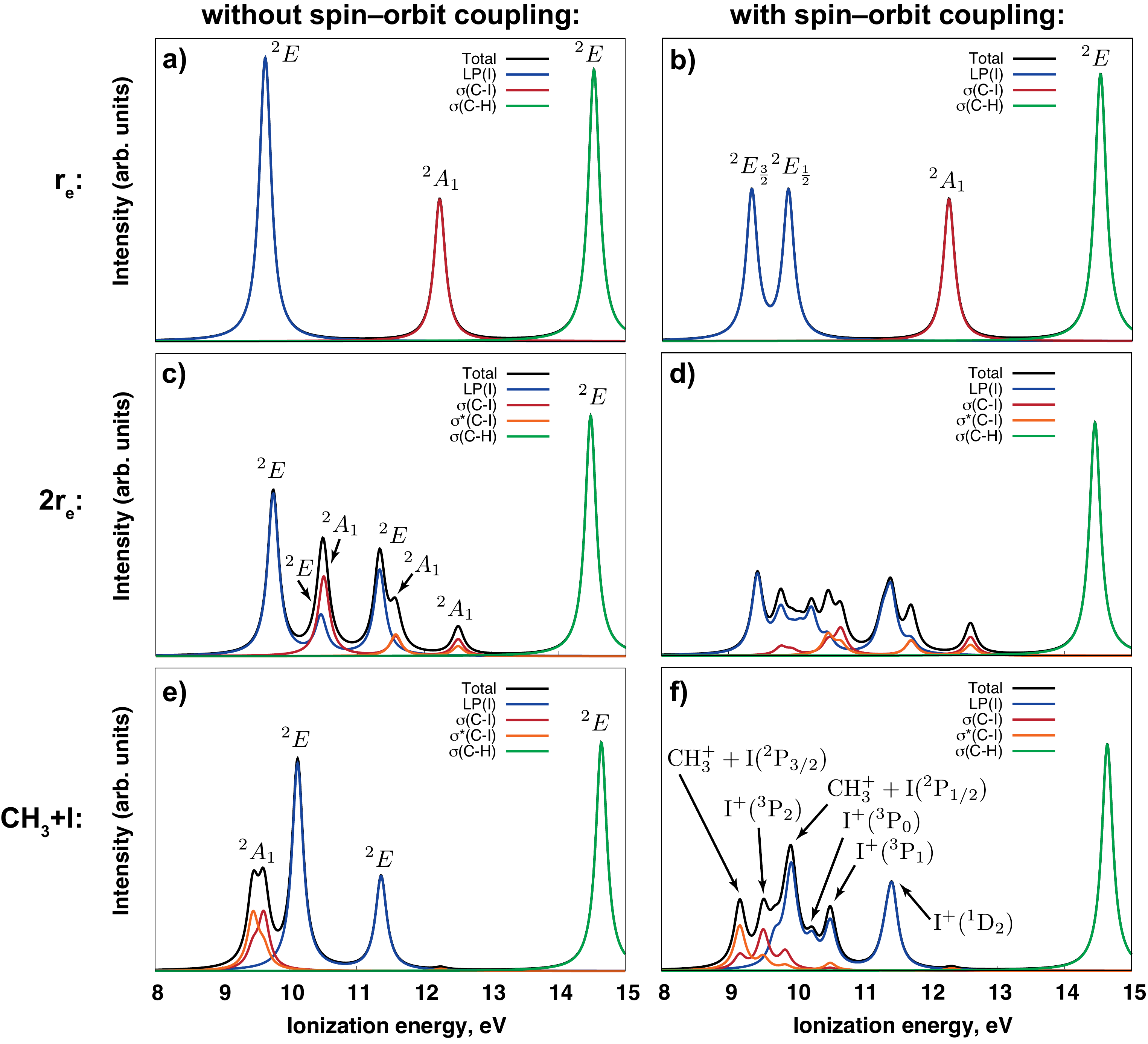} 
	\captionsetup{justification=justified,singlelinecheck=false,font=footnotesize}
	\caption{Photoelectron spectra of methyl iodide (\ce{CH3I}) computed using the DKH2-IP-MR-ADC method at the equilibrium ($r_e$, a and b), stretched ($2r_e$, c and d), and dissociated (\ce{CH3}+I, e and f) geometries. 
	Spectra were calculated with (b, d, f) and without (a, c, e) spin--orbit coupling effects.
	Each plot shows photoelectron intensity contributions from the iodine lone-pair (LP(I)), C--I $\sigma$-bonding ($\sigma$(C--I)), C--I  $\sigma$-antibonding ($\sigma^*$(C--I)), and C--H $\sigma$-bonding  ($\sigma$(C--H)) orbitals.
	} 
	\label{fig:ch3i}
\end{figure*}

Finally, we showcase the multireference capabilities of our spin--orbit EA/IP-ADC implementation by simulating the photoelectron spectrum of methyl iodide (\ce{CH3I}) along the C--I bond dissociation.
Due to its small size, dissociative low-lying excited states, and strong spin--orbit coupling, \ce{CH3I} has become a prototype for testing new experimental and theoretical techniques aimed at understanding the electronic structure and coupled electron-nuclear dynamics at atto- and femtosecond times scales.\cite{Ragle1970:p178184,Woodward1986:p274278,Dobber1993:p836853,Fahr1995:p195203,Schultz1997:p50315034,Olney1998:p211237,Urban2002:p49384947,Hu2007:p68136821,Alekseyev2007:p234102,Alekseyev2007:p234103,DeNalda2008:p244309,Locht2009:p109128,Dura2009:p134311,Rubio-Lago2009:p174309,Locht2010:p105101,Thire2010:p1564415652,Kartakoullis2013:p2238322390,MarggiPoullain2017:p78867896,Forbes2018:p144302,Forbes2018:p94304,Warne2020:p2569525703,Downes-Ward2021:p134003}
Most studies have focused on investigating the \ce{CH3I} photodissociation dynamics following an excitation into the first absorption band at 220 -- 350 nm (so-called $A$-band), which promotes electrons from the iodine lone pairs into the C--I antibonding orbital ($n$ $\rightarrow$ $\sigma^*$). 
In particular, time-resolved (pump-probe) photoelectron spectroscopy provided valuable insights about the \ce{CH3I} photodissociation mechanism by measuring electron binding energies as a function of time.\cite{Dobber1993:p836853,Schultz1997:p50315034,Hu2007:p68136821,Dura2009:p134311,Forbes2018:p144302,Thire2010:p1564415652,Warne2020:p2569525703,Downes-Ward2021:p134003}
Comparing the results of these measurements with accurate theoretical calculations provides opportunities to obtain deeper insights about the interplay of spin--orbit coupling, strong electron correlation, and nonadiabatic relaxation in photodissociation dynamics.

Here, we investigate the effect of spin--orbit coupling on the photoelectron spectra of \ce{CH3I} computed at equilibrium ($r_e$), stretched ($2r_e$), and completely dissociated (\ce{CH3}+I) geometries.
In the stretched structure, the C--I bond was elongated by a factor of two relative to its equilibrium value but the geometry of \ce{CH3} fragment was kept frozen.
For the dissociated structure, the iodine atom was placed $\sim$ $6.7$ \AA\ away from the carbon atom and the geometry of \ce{CH3} moiety was fully optimized.

\cref{fig:ch3i} shows the $r_e$, $2r_e$, and \ce{CH3}+I photoelectron spectra simulated using DKH2-IP-MR-ADC(2)-X with and without spin--orbit coupling effects.
The $r_e$ photoelectron spectrum simulated without spin--orbit coupling (\cref{fig:ch3i}a) exhibits only three peaks corresponding to the electron detachment from the iodine lone pairs ($1^2E$, LP(I)), the  C--I $\sigma$-bonding orbital ($1^2A_1$, $\sigma$(C--I)), and the  C--H bonding orbitals of \ce{CH3} fragment ($2^2E$, $\sigma$(C--H)).
Including spin--orbit coupling splits the $1^2E$ transition into the $1^2E_{3/2}$ and $1^2E_{1/2}$ components with the zero-field splitting (ZFS) of 0.55 eV at the DKH2-IP-MR-ADC(2)-X/X2C-TZVPall level of theory (\cref{fig:ch3i}b).
The computed ($1^2E_{3/2}$; $1^2E_{1/2}$) vertical ionization energies (9.33; 9.88 eV) are in a good agreement with the experimental binding energies (9.54; 10.17 eV) reported by Locht et al.\cite{Locht2010:p105101}
For the two higher-lying states ($1^2A_1$, $2^2E$), the experimental photoelectron spectrum shows broad bands at 12.1-13.1 and 14-15.6 eV with maxima at 12.6 and 14.8 eV.
These measurements agree well with the calculated ($1^2A_1$; $2^2E_{3/2}$; $2^2E_{1/2}$) vertical ionization energies of (12.27; 14.52; 14.53) eV where the $2^2E_{3/2}$ -- $2^2E_{1/2}$ splitting ($\sim$ 90 \cm $\approx$ 0.01 eV) is due to a very weak spin--orbit coupling in the ionized \ce{CH3} group.
It is important to point out that the $2^2E$ states correspond to ionizing the non-active $8e$ molecular orbitals.
Since DKH2-IP-MR-ADC(2)-X incorporates the full spectrum of single and double excitations (\cref{sec:theory:nonrel_mradc}), the $2^2E$  transitions can be included without expanding the active space. 

Stretching the C--I bond by a factor of two ($2r_e$) results in a more complicated photoelectron spectrum. 
Comparing the $r_e$ and $2r_e$ spectra without spin--orbit coupling effects (Figures \ref{fig:ch3i}a and \ref{fig:ch3i}c), large red shift and lowering of intensity are observed for the lowest-energy $^2A_1$ peak due to the weakening of $\sigma$(C--I).
In addition, two new $^2A_1$ signals appear with smaller intensities.
As shown in \cref{fig:ch3i}c, these features correspond to the ionization of C--I antibonding orbital ($\sigma^*$(C--I)) that is significantly populated at this stretched geometry. 
Since the  $^2E$ state is localized on iodine lone pairs (LP(I)), its energy increases by only 0.13 eV.
However, a significant fraction of $^2E$  intensity is transferred into the higher-lying $^2E$ states that appear 0.7 and 1.6 eV higher in energy.
Incorporating spin--orbit coupling results in the zero-field splitting of $^2E$ states and allows them to interact with $^2A_1$, which further complicates the spectrum (\cref{fig:ch3i}d).
Although we cannot assign symmetries for each peak in \cref{fig:ch3i}d, we note that the energy separations and orbital character of states in our DKH2-IP-MR-ADC(2)-X calculations with and without spin--orbit coupling are in a good agreement with the results of a multireference configuration interaction study by Marggi Poullain and co-workers.\cite{MarggiPoullain2017:p78867896}
Interestingly, incorporating spin--orbit coupling results in a much stronger overlap of photoelectron signals from $\sigma$(C--I) and $\sigma^*$(C--I), which indicates that this effect facilitates bond breaking at this geometry.

Finally, we consider the photoelectron spectra computed for the fully dissociated \ce{CH3}+I structure with a relaxed (planar) \ce{CH3} fragment.
Without spin--orbit effects (\cref{fig:ch3i}e), the \ce{CH3}+I spectrum exhibits fewer features compared to that at the $2r_e$ geometry (\cref{fig:ch3i}c).
Relaxing the \ce{CH3} geometry red shifts the two lowest-energy $^2A_1$ transitions corresponding to the ionization of \ce{CH3} radical and I atom.
As a result of complete C--I bond dissociation, the first $^2E$ transition blue shifts by $\sim$ 0.37 eV, gaining intensity relative to the $2r_e$ spectrum. 
Incorporating spin--orbit coupling (\cref{fig:ch3i}f) significantly perturbs the spectrum, splitting the peaks and allowing the resulting states interact. 
As discussed in Ref.\@ \citenum{MarggiPoullain2017:p78867896}, the ionized states of \ce{CH3}+I can be assigned to the \ce{CH3} + \ce{I+} and \ce{CH3+} + \ce{I} dissociation limits with \ce{I} or \ce{I+} in their ground or excited electronic states.
Due to spatial symmetry breaking in the state-specific reference CASSCF wavefunction, the degeneracy of some \ce{CH3} + \ce{I+} and \ce{CH3+} + \ce{I} states in our calculations is lifted by $\sim$ 0.05 eV on average with a maximum of $\sim$ 0.15 eV.
Despite this, for the features with significant intensity tentative assignments can be made as follows: \ce{CH3+} + \ce{I}($^2P_{3/2}$) [9.2 eV], \ce{CH3} + \ce{I+}($^3P_{2}$) [9.5 eV], \ce{CH3+} + \ce{I}($^2P_{1/2}$) [9.9 eV], \ce{CH3} + \ce{I+}($^3P_{0}$) [10.3 eV], \ce{CH3} + \ce{I+}($^3P_{1}$) [10.5 eV], and \ce{CH3} + \ce{I+}($^1D_{2}$) [11.4 eV].
For the \ce{CH3+} + \ce{I} ionization channel, these results are in a good agreement with the data from femtosecond pump-probe experiments by de Nalda et al.\cite{DeNalda2008:p244309} that reported the first ionization energy of $\sim$ 9.3 eV and the \ce{I}($^2P_{3/2}$) -- \ce{I}($^2P_{1/2}$) zero-field splitting of $\sim$ 0.8 eV.
In the \ce{CH3} + \ce{I+} channel, the energy separations of \ce{I+} levels ($^3P_{0}$ -- $^3P_{2}$, $^3P_{1}$ -- $^3P_{0}$, $^1D_{2}$ -- $^3P_{1}$) computed using DKH2-IP-MR-ADC(2)-X (0.8, 0.2, 0.9 eV) agree well with the data from atomic spectroscopy (0.8, 0.1, 0.8 eV).\cite{NIST_ASD}

\section{Conclusion}
\label{sec:conclusions}
We presented a formulation of algebraic diagrammatic construction theory that enables simulating charged electronic states and photoelectron spectra with a computationally efficient treatment of electron correlation (both static and dynamic) and spin--orbit coupling.
Starting with either a restricted Hartree--Fock or a complete active space self-consistent field reference wavefunction, our implementation allows to perform single-reference (SR-) or multireference (SR- and MR-) ADC calculations incorporating dynamic correlation and spin--orbit coupling up to the second order in perturbation theory.
The relativistic effects are described using three flavors of two-component spin--orbit Hamiltonians, namely: Breit--Pauli, exact two-component first-order Douglas--Kroll--Hess (sf-X2C+so-DKH1), and exact two-component second-order Douglas--Kroll--Hess (sf-X2C+so-DKH2).

We benchmarked the accuracy of spin--orbit SR- and MR-ADC methods for simulating zero-field splitting and photoelectron spectra of atoms and small molecules.
When multireference effects are not important, such as in main group atoms and diatomics, the SR-ADC methods are competitive in accuracy to the MR-ADC approximations, often showing better agreement with experimental results. 
However, as we demonstrated in our studies of $d^9$ transition metal atoms and the methyl iodide molecule, the MR-ADC methods are more reliable in excited states and can correctly describe photoelectron spectra in non-equilibrium regions of potential energy surfaces that can be important for interpreting the results of time-resolved experiments.

Overall, our benchmark results demonstrate that the ADC methods developed in this work are promising techniques for efficient and accurate simulations of spin--orbit coupling in charged electronic states.
To make them practical, several developments are still necessary, such as efficient computer implementation, enabling calculations for degenerate or state-averaged reference states, and extensions to neutral excitations.
The spin--orbit ADC methods are also attractive for simulating how matter interacts with high-energy light, as was demonstrated in a recent study of time-resolved X-ray photoelectron spectra along iron pentacarbonyl photodissociation.\cite{Gaba2024:p1592715938}.
Pushing these frontiers holds promise for improving our understanding of relativistic effects and electron correlation in increasingly complicated molecular systems.

\section{Appendix: Deriving Amplitude Equations for the Internal Single Excitations}

As discussed in \cref{sec:theory:soc_mradc}, incorporating ${H}_{\mathrm{SO}}$ (\cref{eq:h_2c_so_general}) in the perturbation term $V$ of MR-ADC effective Hamiltonian (\cref{eq:eff_H1,eq:eff_H2}) results in new contributions to $\mathbf{M}_{\pm}$ (\cref{eq:Mmatrix_ea,eq:Mmatrix_ip}) starting at the first order in perturbation theory.
Since ${H}_{\mathrm{SO}}$ contains terms with all active indices (i.e., spin--orbit coupling in active orbitals), diagonal blocks of $\mathbf{M}^{(k)}_{\pm}$ ($k\ge1$) with the excitation operators $h^{(l)\dagger}_{\pm\mu}$ and $h^{(l)}_{\pm\nu}$  belonging to the same class will get modified.
As an example, we consider the diagonal sectors of $\mathbf{M}^{(k)}_{+}$ in spin--orbit EA-MR-ADC that can be written as:
\begin{align}
		M^{(k)}_{+\mu\nu} 
		&= \sum^{l+m=k}_{lm} \langle{\Psi_0}|[h^{(l)}_{+\mu} , [\tilde{H}^{(m)}, h^{(l)\dagger}_{+\nu} ]]_{+}|{\Psi_0}\rangle \notag \\
		\label{eq:m_explanded_1}
		&= \sum^{l+m=k}_{lm} \left( \langle{\Psi_0}|h^{(l)}_{+\mu}  \tilde{H}^{(m)} h^{(l)\dagger}_{+\nu} |{\Psi_0}\rangle \right. \notag\\
			&-\left.\langle{\Psi_0}|h^{(l)}_{+\mu} h^{(l)\dagger}_{+\nu} \tilde{H}^{(m)}  |{\Psi_0}\rangle \right)
\end{align}
We also write down an expression for the same diagonal block of $\mathbf{M}^{(k)\dag}_{\pm}$:
\begin{align}
	M^{(k)\dag}_{+\mu\nu} 
	&= \sum^{l+m=k}_{lm} \langle{\Psi_0}|[h^{(l)}_{+\nu} , [\tilde{H}^{(m)}, h^{(l)\dagger}_{+\mu} ]]_{+}|{\Psi_0}\rangle^\dag \notag \\
	\label{eq:m_explanded_2}
	&= \sum^{l+m=k}_{lm} \left( \langle{\Psi_0}|h^{(l)}_{+\mu}  \tilde{H}^{(m)} h^{(l)\dagger}_{+\nu} |{\Psi_0}\rangle \right. \notag\\
	&-\left.\langle{\Psi_0}|\tilde{H}^{(m)}   h^{(l)}_{+\mu} h^{(l)\dag}_{+\nu} |{\Psi_0}\rangle \right)
\end{align}
where we used the fact that $\tilde{H}^{(m)}$ is Hermitian at any order $m$. 
Comparing \cref{eq:m_explanded_1,eq:m_explanded_2}, we note that for the effective Hamiltonian matrix to be Hermitian ($M^{(k)}_{+\mu\nu} = M^{(k)\dag}_{+\mu\nu}$) their last terms should be zero or equal to each other.
Since $h^{(l)}_{+\mu}$ and $h^{(l)\dagger}_{+\nu}$  are from the same class, these contributions correspond to the projections of $\tilde{H}^{(m)}$ by excitations inside active space (so-called internal excitations).
Due to the all-active contributions from ${H}_{\mathrm{SO}}$, the last two terms in the \cref{eq:m_explanded_1,eq:m_explanded_2} are generally not the same, unless the effective Hamiltonian is parameterized to prevent that.

To ensure that $\mathbf{M}^{(k)}_{\pm}$ is rigorously Hermitian up to $k = 2$, we incorporate a new class of first-order internal excitations in the correlation operator $T$:
\begin{align}
	T^{(1)} &\leftarrow \sum_{x>y} t^{y(1)}_x a^{\dagger}_y a_x
\end{align} 
which ensure that the last two terms of  \cref{eq:m_explanded_1,eq:m_explanded_2} (and similar terms in IP-MR-ADC) are equal to each other.\cite{Mazin2021:p61526165}
The $t^{y(1)}_x$ ($x>y$) amplitudes are determined by solving a system of linear equations:
\begin{align}
	\label{eq:app_amp_int}
	\langle{\Psi_0}|a^{\dagger}_x a_y \tilde{H}^{(1)}|{\Psi_0}\rangle - \langle{\Psi_0}|a^{\dagger}_y a_x \tilde{H}^{(1)} |{\Psi_0}\rangle^{*} = 0
\end{align} 
Since $\tilde{H}^{(1)}$ depends on the complex-valued perturbation operator $V_{\mathrm{2c}}$, \cref{eq:app_amp_int} needs to be solved for $\mathrm{Re}{(t^{y(1)}_x)}$ and $\mathrm{Im}{(t^{y(1)}_x)}$ separately.
Each system of equations can be written in a tensor form:
\begin{align}
\label{eq:real_amp_proj}
\mathbf{K}^{}_{\mathrm{Re}}\mathbf{T}^{(1)}_{\mathrm{Re}} &= -\mathbf{V}^{}_{\mathrm{Re}} \\
\label{eq:imag_amp_proj}
\mathbf{K}^{}_{\mathrm{Im}}\mathbf{T}^{(1)}_{\mathrm{Im}} &= -\mathbf{V}^{}_{\mathrm{Im}}
\end{align}
where $\mathbf{T}^{(1)}_{\mathrm{Re}}$ and $\mathbf{T}^{(1)}_{\mathrm{Im}}$ contain the real and imaginary parts of $t^{y(1)}_x$ ($x>y$), respectively. 
The elements of $\mathbf{K}^{}_{\mathrm{Re}}$, $\mathbf{K}^{}_{\mathrm{Im}}$, $\mathbf{V}^{}_{\mathrm{Re}}$, and $\mathbf{V}^{}_{\mathrm{Im}}$ are defined as:
\begin{align}
K^{\mathrm{Re}}_{xy,wz} &= \langle{\Psi_0}|(a^{\dagger}_x a_y - a^{\dagger}_y a_x) [H^{(0)}, a^{\dagger}_z a_w - a^{\dagger}_w a_z] |{\Psi_0}\rangle \\
K^{\mathrm{Im}}_{xy,wz} &= \langle{\Psi_0}|(a^{\dagger}_x a_y + a^{\dagger}_y a_x) [H^{(0)}, a^{\dagger}_z a_w + a^{\dagger}_w a_z] |{\Psi_0}\rangle \\
V^{\mathrm{Re}}_{xy} &= \mathrm{Re}(\langle{\Psi_0}| (a^{\dagger}_x a_y - a^{\dagger}_y a_x) V_{\mathrm{2c}} |{\Psi_0}\rangle) \\
V^{\mathrm{Im}}_{xy}  &= \mathrm{Im}(\langle{\Psi_0}| (a^{\dagger}_x a_y + a^{\dagger}_y a_x) V_{\mathrm{2c}} |{\Psi_0}\rangle)
\end{align}
where $H^{(0)}$ is the Dyall zeroth-order Hamiltonian and $V_{\mathrm{2c}}$ is the perturbation operator defined in \cref{eq:v_2c}.

To solve \cref{eq:real_amp_proj,eq:imag_amp_proj}, we first diagonalize the real-valued and Hermitian  $\mathbf{K}^{}_{\mathrm{Re}}$ and $\mathbf{K}^{}_{\mathrm{Im}}$ matrices:
\begin{align}
	\label{eq:gen_eig_kint}
	\mathbf{K}^{}_{\mathrm{Re}} \mathbf{Z}^{}_{\mathrm{Re}} &= \mathbf{S}^{}_{\mathrm{Re}} \mathbf{Z}^{}_{\mathrm{Re}} \boldsymbol{\epsilon}^{}_{\mathrm{Re}} \\
	\label{eq:gen_eig_kprimeint}
	\mathbf{K}^{}_{\mathrm{Im}} \mathbf{Z}^{}_{\mathrm{Im}} &= \mathbf{S}^{}_{\mathrm{Im}} \mathbf{Z}^{}_{\mathrm{Im}} \boldsymbol{\epsilon}^{}_{\mathrm{Im}}
\end{align}
where $\mathbf{Z}^{}_{\mathrm{Re}}$ and $\mathbf{Z}^{}_{\mathrm{Im}}$ denote the eigenvectors of corresponding generalized eigenvalue problems and $\mathbf{S}^{}_{\mathrm{Re}}$ and $\mathbf{S}^{}_{\mathrm{Im}}$ are the overlap matrices:
\begin{align}
	S^{\mathrm{Re}}_{xy,wz} &= \langle{\Psi_0}|(a^{\dagger}_x a_y - a^{\dagger}_y a_x)(a^{\dagger}_z a_w - a^{\dagger}_w a_z)|{\Psi_0}\rangle \\
	S^{\mathrm{Im}}_{xy,wz} &= \langle{\Psi_0}|(a^{\dagger}_x a_y + a^{\dagger}_y a_x)(a^{\dagger}_z a_w + a^{\dagger}_w a_z)|{\Psi_0}\rangle
\end{align}
The contributions to internal amplitudes can then be obtained as follows:
\begin{align}
\mathbf{T}^{(1)}_{\mathrm{Re}} &= -\mathbf{S}_{\mathrm{Re}}^{-1/2}\mathbf{\tilde{Z}}^{}_{\mathrm{Re}}\boldsymbol{\epsilon}_{\mathrm{Re}}^{-1}\mathbf{\tilde{Z}}_{\mathrm{Re}}^{\dagger}\mathbf{S}_{\mathrm{Re}}^{-1/2}\mathbf{V}_{\mathrm{Re}} \\
\mathbf{T}^{(1)}_{\mathrm{Im}} &= -\mathbf{S}_{\mathrm{Im}}^{-1/2}\mathbf{\tilde{Z}}_{\mathrm{Im}}^{}\boldsymbol{\epsilon}_{\mathrm{Im}}^{ -1}\mathbf{\tilde{Z}}_{\mathrm{Im}}^{ \dagger}\mathbf{S}_{\mathrm{Im}}^{ -1/2}\mathbf{V}_{\mathrm{Im}}^{} 
\end{align}
where $\mathbf{\tilde{Z}}^{}_{\mathrm{Re}} = \mathbf{S}^{1/2}_{\mathrm{Re}}\mathbf{Z}^{}_{\mathrm{Re}}$ and $\mathbf{\tilde{Z}}^{}_{\mathrm{Im}} = \mathbf{S}^{1/2}_{\mathrm{Im}}\mathbf{Z}^{}_{\mathrm{Im}}$.

\acknowledgement
This work was supported by the National Science Foundation, under Grant No.\@ CHE-2044648.
Computations were performed at the Ohio Supercomputer Center under Project No.\@ PAS1583.\cite{OhioSupercomputerCenter1987}

\suppinfo
Additional computational details, including geometries, reference active spaces, and the selection of CASCI states for each calculation.


\providecommand{\latin}[1]{#1}
\makeatletter
\providecommand{\doi}
  {\begingroup\let\do\@makeother\dospecials
  \catcode`\{=1 \catcode`\}=2 \doi@aux}
\providecommand{\doi@aux}[1]{\endgroup\texttt{#1}}
\makeatother
\providecommand*\mcitethebibliography{\thebibliography}
\csname @ifundefined\endcsname{endmcitethebibliography}
  {\let\endmcitethebibliography\endthebibliography}{}

\end{document}